\documentclass[twocolumn]{aastex631}

\newcommand{\fluxcgs}{\ensuremath{\mathrm{erg}\, \mathrm{s}^{-1}\, \mathrm{cm}^{-2}}}
\newcommand{\phflux}{\ensuremath{\mathrm{ph}\, \mathrm{s}^{-1}\, \mathrm{cm}^{-2}}}

\newcommand{\gm}{$\gamma$}
\accepted{\today}

\submitjournal{ApJ}

\shorttitle{DA 362}
\shortauthors{Swain et al.}

\begin{document}

\title{DA 362: A Gamma-ray Emitting Compact Symmetric Object}

\correspondingauthor{Subhashree Swain, Vaidehi S. Paliya}
\email{subhashree.swain@iucaa.in, vaidehi.s.paliya@gmail.com}
\author[0009-0006-2098-2592]{Subhashree Swain}
\affiliation{Inter-University Centre for Astronomy and Astrophysics (IUCAA), SPPU Campus, Pune 411007, India}
\author[0000-0001-7774-5308]{Vaidehi S. Paliya}
\affiliation{Inter-University Centre for Astronomy and Astrophysics (IUCAA), SPPU Campus, Pune 411007, India}
\author[0000-0002-4464-8023]{D. J. Saikia}
\affiliation{Inter-University Centre for Astronomy and Astrophysics (IUCAA), SPPU Campus, Pune 411007, India}
\author[0000-0002-4464-8023]{C. S. Stalin}
\affiliation{Indian Institute of Astrophysics, Block II, Koramangala, Bengaluru 560034, Karnataka, India}

\begin{abstract}

The \gm-ray detection from an astrophysical object indicates the presence of an extreme environment where high-energy radiation is produced. With the continuous monitoring of the \gm-ray sky by the Fermi Large Area Telescope (LAT), leading to deeper sensitivity, the high-energy \gm-ray emission has now been detected from a diverse class of jetted active galactic nuclei (AGN). Here, we present the results of a multiwavelength study of the radio source DA~362, which was reported to be a blazar candidate of uncertain type. However, it was recently identified as a bona fide compact symmetric object (CSO) based on its sub-kpc, bi-polar radio morphology, and lack of radio variability. This makes DA~362 the only fourth \gm-ray emitting object of this enigmatic class of radio-loud AGN. Using five very long baseline interferometry observations covering 1996-2018, we found the jet separation velocity to be subluminal ($v_{\rm app}\sim 0.2c$), thus supporting its CSO nature. Its Fermi-LAT observations revealed a \gm-ray flaring activity, a phenomenon never detected from the other three \gm-ray detected CSOs. This object is bright in the near-infrared band but extremely faint in the optical-ultraviolet filters, hinting at possible obscuration. The Swift X-Ray Telescope observation of DA 362 reveals an extremely hard X-ray spectrum, though a strong claim cannot be made due to large uncertainties. We conclude that deeper observations are needed to probe the broadband properties of this enigmatic object and to understand the origin of high-energy \gm-ray emission.

\end{abstract}


\section{Introduction} \label{sec:intro}
Compact Symmetric Objects (CSOs) are a special class of active galactic nuclei (AGN) hosting sub-kiloparsec-scale jets and exhibiting symmetric radio morphologies \citep[e.g.,][]{2016AN....337....9O}. They are thought to host misaligned jets, and hence, the observed broadband emission is not expected to be beamed \citep[cf.][]{Wilkinson1994, readhead1996}. These enigmatic objects are likely to be in the early stage of their evolution with kinematic ages smaller than a few thousand years \citep[e.g.,][]{1996ApJ...460..612R}. The small sizes CSOs could be due to their young age, dense galactic environment inhibiting the jet propagation, and/or recurrent/transient episodes of the nuclear jet activity \cite[see,][for a review]{2021A&ARv..29....3O}. From the analysis of their relative numbers, and redshift and linear size distributions, \citet[][]{2024ApJ...961..241K} presented strong evidence that CSOs do not evolve into larger-scale radio sources and that they should be considered as a distinct `short-lived' jetted AGN population rather than `young' AGN \citep[see also,][]{2024ApJ...961..242R}.

The number of \gm-ray emitting CSOs remained tiny compared to more common Fanaroff-Riley Type I and II misaligned jetted AGN and blazars. Only three CSOs are reported to be detected with the Fermi Large Area Telescope (LAT): TXS 0128+554 \citep[$z=0.036$; associated with the \gm-ray source 4FGL J0131.2+5547;][]{lister2020}, NGC 3894 \citep[$z=0.012$; counterpart of 4FGL J1149.0+5924;][]{Principe2020}, and NGC 6328 \citep[$z=0.015$; associated with 4FGL J1724.2$-$6501;][]{2016ApJ...821L..31M}. All of them are located in the nearby Universe ($z<0.05$), and their proximity could be the primary reason for their Fermi-LAT detection if the \gm-ray emission is unbeamed \citep[e.g.,][]{2021MNRAS.507.4564P}. The \gm~rays can be produced by the hadronic mechanisms or due to the interaction of the relativistic electrons present in the lobes with the low-energy optical-UV photons originating from the accretion disk \citep[cf.][]{Stawarz2008,2011MNRAS.412L..20K}. The radio lobes typically expand with sub-relativistic velocities; therefore, the \gm-ray radiation is not expected to show significant flux variations, especially on short timescales ($\sim$weeks-to-months). Indeed, none of the three \gm-ray detected CSOs have exhibited significant flux variability as of now \citep[][]{2023ApJS..265...31A}. However, a definite conclusion cannot be drawn given the small number of known \gm-ray emitting CSOs. Increasing the sample size of these peculiar sources is also crucial to investigate the radiative processes powering their jets and their interaction with the surrounding environment, compare them with other non-\gm-ray detected CSOs, and understand their evolution.

Recently, \citet[][]{2024ApJ...961..240K} presented a comprehensive catalog of 79 CSOs searched from the literature and by analyzing their multi-frequency radio observations. They adopted the following four criteria: (i) projected jet length $<$1 kpc, (ii) detection of bi-polar radio emission, (iii) non-variable nature, and (iv) no superluminal motion detection in excess of $v_{\rm app}=2.5c$. In order to identify the potential \gm-ray emitting CSOs, we cross-matched the CSO catalog with the fourth data release of the fourth \gm-ray source catalog of Fermi-LAT detected objects \citep[4FGL-DR4;][]{2020ApJS..247...33A,ballet2023}. Using a search radius of 5$^{\prime\prime}$, this exercise led to the identification of four \gm-ray sources\footnote{In the sample of \citet[][]{2024ApJ...961..240K}, there is a CSO, B3 0822+394, whose \gm-ray detection was claimed by \citet[][]{2022ApJ...927..221G}. However, this object is not reported as a \gm-ray source in the 4FGL-DR4 catalog. Therefore, it did not appear in our final sample of \gm-ray emitting CSOs.}. Among them, three are already reported as \gm-ray emitting CSOs in earlier works \citep[][]{2016ApJ...821L..31M,Principe2020,lister2020}. The fourth object, DA~362 (B2~1413+34), which is associated with the \gm-ray source 4FGL~J1416.0+3443, is classified as a blazar candidate of uncertain type in the 4FGL-DR4 catalog. However, as shown by \citet[][]{2024ApJ...961..240K}, this object is a bona fide CSO, thereby making it only the fourth \gm-ray detected object of this class. Moreover, \citet[][]{2021ApJS..256...13B} reported the detection of a transient \gm-ray elevated activity from this object. In this article, we present the results of an investigation of its multi-wavelength properties utilizing $>$15 years of the Fermi-LAT data and other low-frequency observations and compare them with other \gm-ray detected CSOs. Section~\ref{sec:data} elaborates on the data reduction steps. The results are presented in Section~\ref{sec:res} and discussed in Section~\ref{sec:dis}. We summarize our findings in Section~\ref{sec:sum}. Throughout, we adopt the convention $S_{\nu}$ $\sim$ $\nu^{\alpha}$ for spectral index $\alpha$, and use the cosmological parameters $\Omega_m$=0.27, $\Omega_{\bigwedge}$=0.73 and H$_o$=70 km s$^{-1}$ Mpc$^{-1}$.

\section{Data Reduction and Analysis}\label{sec:data}
\subsection{Fermi-LAT}\label{fermi}
We followed the standard data reduction procedure to analyze the Fermi-LAT Pass 8 data of DA~362 covering 2008 August 4 to 2024 April 5 (MJD 54683$-$60405). We considered SOURCE class events in the energy range 0.1$-$300 GeV and lying within a region of interest of radius 15$^{\circ}$ centered at the target AGN. To avoid contamination from Earth's albedo, a zenith angle cut of $z_{\rm max}<90^{\circ}$ was also applied. All 4FGL-DR4 cataloged sources located within 20$^{\circ}$ of DA~362 position were considered to model the \gm-ray sky in the likelihood fitting. To take into account the diffuse background emission, we adopted the Galactic and isotropic background templates provided by the Fermi Science Support Center\footnote{\url{https://fermi.gsfc.nasa.gov/ssc/data/access/lat/BackgroundModels.html}}. The \gm-ray spectral parameters of all sources were first optimized, and then the final likelihood fitting was performed by varying the parameters of all sources with detection significance $>5\sigma$ \citep[test statistic, TS$>$25;][]{1996ApJ...461..396M}.

We also generated the \gm-ray spectrum (in six energy bins covering 0.1$-$300 GeV) and monthly-binned light curve of DA~362 with the same settings described above. In the time/energy bins of non-detections (TS$<$9), we computed flux upper limits at 95\% confidence level. 

\subsection{Swift}\label{swift}
The Neil Gehrels Swift satellite observed DA~362 on 2023 March 31 and 2023 April 5 (target id: 15938, PI: Paliya). The Swift X-Ray Telescope (XRT) data were analyzed using the online Swift-XRT data products generator\footnote{\url{https://www.swift.ac.uk/user\_objects/}} \citep[][]{2009MNRAS.397.1177E}. Since there were no previous X-ray observations of the source, we first checked its X-ray detection. An X-ray source was found at the right ascension (RA) of 14h 16m 04.03s and declination (Dec) of +34$^{\circ}$ 44$^{\prime}$ 34.8$^{\prime\prime}$ with 90\% confidence radius of 5.7 arcseconds. The angular separation between the optimized X-ray and radio positions was estimated to be 2.5 arcseconds, thus confirming that the X-ray source is spatially consistent with DA~362. Given the low-exposure individual pointings, we added both observations to generate a combined X-ray spectrum with a net exposure of 4.2 ksec in which a net 22 counts were detected from the source. The obtained spectrum was grouped using the task {\tt grppha} to have at least one count per bin, and the C-statistic was employed for the spectral fitting in XSPEC \citep[][]{1996ASPC..101...17A}. Uncertainties were estimated at 90\% confidence level. For the plotting purpose, the fitted X-ray spectrum was re-binned to have at least 3$\sigma$ detection in each bin or grouped in sets of 3 bins.

The Swift UltraViolet and Optical Telescope (UVOT) data were reduced following the recommended guidelines. DA~362 was observed in two UV filters, namely UVW1 and UVM2. We combined the individual frames using the task {\tt uvotimsum}. For the photometry, we applied the command {\tt uvotsource}. The source was undetected, with UVM2 and UVW1 magnitudes fainter than 19.21 and 21.01 (Vega system), respectively.

\subsection{Radio Observations}
The five epochs of Very Large Baseline Interferometry (VLBI) observations of DA 362 covering 22 years (1996 to 2018) are provided on the Astrogeo website\footnote{http://astrogeo.org} \citep[][]{2021AJ....161...14P}. Four observations were in the X and S bands, and one epoch of data was taken in the C band. Additionally, the source was also detected in the Low-Frequency Array Two-metre Sky Survey, Very Large Array Sky Survey, and Rapid ASKAP Continuum Survey \citep[][]{2020PASP..132c5001L,2022A&A...659A...1S,2024PASA...41....3D}.

\subsection{Other Observations}
We collected archival flux measurements from Space Science Data Center\footnote{\url{https://tools.ssdc.asi.it/SED/}}. These datasets were published in several radio-to-optical/UV catalogs \citep[][]{1984ApJ...278L...1N,1990PKS...C......0W,1994ApJS...91..111W,1994Ap&SS.217....3E,1996ApJS..103..427G,1998AJ....115.1693C,2003MNRAS.342.1117M,2006AJ....131.1163S,2007ApJS..171...61H,2010AJ....140.1868W,2010yCat.2298....0Y,2014Ap&SS.354...97Y,2015ApJ...801...26H,2020ApJS..249....3A}. Since DA 362 was undetected in the Swift-UVOT data analysis, we checked whether it was included in the Panoramic Survey Telescope and Rapid Response System catalog \citep[PanSTARRS;][]{2016arXiv161205560C}. A faint optical source positionally consistent with DA~362 was identified.

\begin{figure}
    \includegraphics[width=\linewidth]{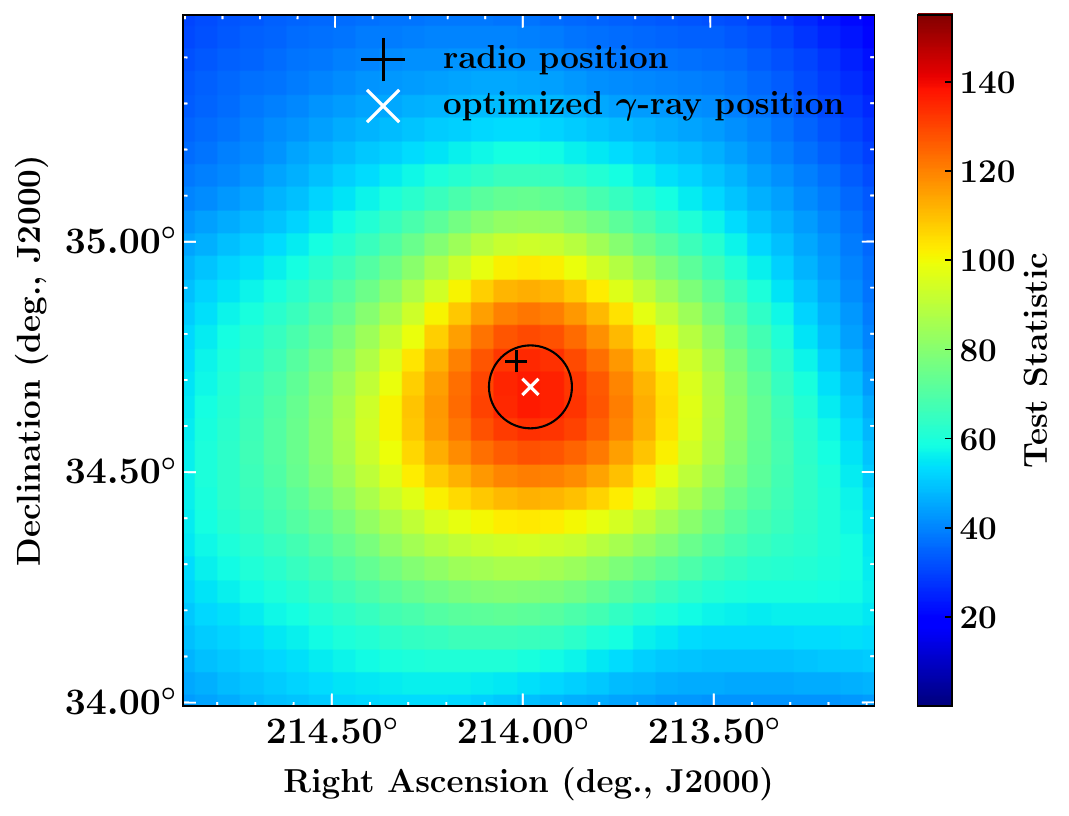}
    \caption{The test statistic map of the \gm-ray sky centered at the optimized \gm-ray position of DA~362. The black circle shows the 95\% uncertainty region for the \gm-ray position. The radio position of DA 362 is highlighted with the `+' mark.}
    \label{fig:1}
\end{figure}

\begin{table*}
\centering
\caption{The \gm- and X-ray spectral parameters of DA~362 and other \gm-ray emitting CSOs as derived from the Fermi-LAT and Swift-XRT data analyses, respectively.}\label{tab:tab0}
\begin{tabular}{lcccc} 
\hline
        &             & Gamma-ray Results       &                            &    \\ 
4FGL Name & Counterpart & $F_{\rm 0.1-300~GeV}$ & $\Gamma_{\rm 0.1-300~GeV}$ & TS \\
          &             & 10$^{-8}$ \phflux     &                            &    \\ 
\hline
J1416.0+3443 & DA 362       & 1.26$\pm$0.07 & 2.71$\pm$0.02 & 166\\
J0131.2+5547 & TXS 0128+554 & 0.56$\pm$0.02 & 2.05$\pm$0.02 & 211\\
J1149.0+5924 & NGC 3894     & 0.32$\pm$0.09 & 2.16$\pm$0.10 & 133\\
J1724.2$-$6501 & NGC 6328   & 0.37$\pm$0.12 & 2.28$\pm$0.12 & 48\\
\hline
    &             & X-ray Results       &                            &    \\ 
    & Exposure & $F_{\rm 0.3-10~keV}$ & $\Gamma_{\rm 0.3-10~keV}$ & C-stat/dof \\
    & ksec      & 10$^{-13}$ \fluxcgs  &                           &            \\
\hline
J1416.0+3443 & 4.2 & 7.25$^{+5.38}_{-3.17}$ & 0.79$^{+0.52}_{-0.46}$  & 22.17/22  \\
J0131.2+5547 & 19.2 & 4.08$^{+0.81}_{-0.69}$ & 3.11$^{+0.37}_{-0.36}$  & 107.94/103  \\
J1149.0+5924 & 5.5 & 1.74$^{+1.41}_{-1.74}$ & 1.57$^{+2.20}_{-3.20}$  & 12.71/11  \\
J1724.2$-$6501 & 35.6 & 7.44$^{+1.34}_{-0.71}$ & 1.66$^{+0.23}_{-0.22}$  & 260.59/263 \\
\hline
\end{tabular}
\end{table*}

\section{Results}\label{sec:res}
Only three CSOs were identified in the \gm-ray band prior to this work, making DA~362 the only fourth \gm-ray emitting object of this class. It is included in the 4FGL-DR4 catalog but was missing from previous Fermi-LAT catalogs. We carried out a dedicated data reduction covering the first $\sim$15.75 years of the Fermi-LAT operation. We also optimized the \gm-ray source position of 4FGL~J1416.0+3443 and estimated it to be RA = 14h 15m 55s and Dec = 34$^{\circ}$ 41$^{\prime}$ 18$^{\prime\prime}$. The 95\% uncertainty in the measured position is 0$^{\circ}$.07. We show the TS map of the \gm-ray region in Figure~\ref{fig:1} where the radio and optimized \gm-ray positions are also overplotted. Within the 95\% uncertainty region, both positions are consistent, thus confirming the association of  DA~362 with the \gm-ray source 4FGL~J1416.0+3443. The spectral parameters obtained from the power law fit are reported in Table 1. For comparison, we also provide the spectral parameters of the other three \gm-ray detected CSOs obtained following the same methodology outlined in Section~\ref{fermi}. The computed \gm-ray spectral parameters are on average consistent with that published in previous works \citep[cf.][]{2021MNRAS.507.4564P}. The minor differences, if any, could be mainly due to the different time periods for the Fermi-LAT data reduction done in our and previous works.

Furthermore, the monthly binned \gm-ray light curve of DA~362 is shown in Figure~\ref{fig:lc}. Though sporadically detected, the source remained mostly in quiescence during the first $\sim$12 years of the Fermi-LAT operation, thus explaining its absence in 4FGL-DR3 and earlier \gm-ray catalogs. Interestingly, a flaring activity was identified during MJD 59075-59287. The brightest \gm-ray flux was found to be ($1.3\pm0.2)\times10^{-7}$ \phflux~which is ten times larger than its mission averaged flux value. The detection significance of the flare peak was found to be $\sim$8$\sigma$ (TS=61).

We fitted the Swift-XRT spectrum of DA~362 with a power law model modified for the Galactic absorption fixed to the neutral hydrogen column density of $N_{\rm H}=1.32\times10^{20}$ cm$^{-2}$ \citep[][]{2005AA...440..775K}. In the energy range of 0.3$-$10 keV, the estimated photon index and energy flux are 0.79$^{+0.52}_{-0.46}$ and 7.25$^{+5.38}_{-3.17}\times10^{-13}$ \fluxcgs, respectively. Given the fact that CSOs are typically misaligned jetted AGN, the observation of such a flat X-ray spectrum indicates a strong absorption of soft X-ray emission usually observed in Compton thick sources \citep[e.g.,][]{2007A&A...466..823G,2018ApJ...854...49M}. Therefore, we attempted adding a redshifted absorption component ({\tt ZTBABS} in XSPEC) during the fit; however, the parameters could not be constrained due to fitting failure. To get an idea about the intrinsic absorption, we froze the photon index to 1.8 and 2, i.e., to a typical value estimated for AGN, during the XSPEC fit. We obtained the fitted $N_{\rm H}$ value to be $1.05^{+1.02}_{-0.63}\times10^{22}$ cm$^{-2}$ and $1.22^{+1.09}_{-0.69}\times10^{22}$ cm$^{-2}$, respectively. These results provide supporting evidence about the possible obscured nature of DA 362. Deeper spectral observations will be needed to characterize the possible X-ray obscured nature of this \gm-ray detected CSO.

DA~362 is extremely faint in the optical-UV band. It remained undetected in the Swift-UVOT observations. PanSTARRS catalog reports a faint optical source with $g$ and $r$-band magnitudes being 21.59$\pm$0.14 and 21.32$\pm$0.18, respectively, positionally consistent with the radio source. However, it remained undetected in the $i$ and $z$ filters. \citet[][]{1992PASA...10..140W} reported its redshift to be $z=0.26$, though its origin, whether spectroscopic or photometric, is unclear. The source lies in the Sloan Digital Sky Survey coverage area; however, it lacks spectroscopic measurement, likely due to its faintness. Moreover, \citet[][]{1993ApJS...88....1S} provided a limit of 24 and 23 magnitudes in the $r$ and $i$ bands, respectively. In contrast, it is well detected in the mid-infrared (MIR) band covered by the Wide-field Infrared Survey Explorer (WISE). 

\begin{figure*}
\hbox{
    \includegraphics[scale=0.48]
    {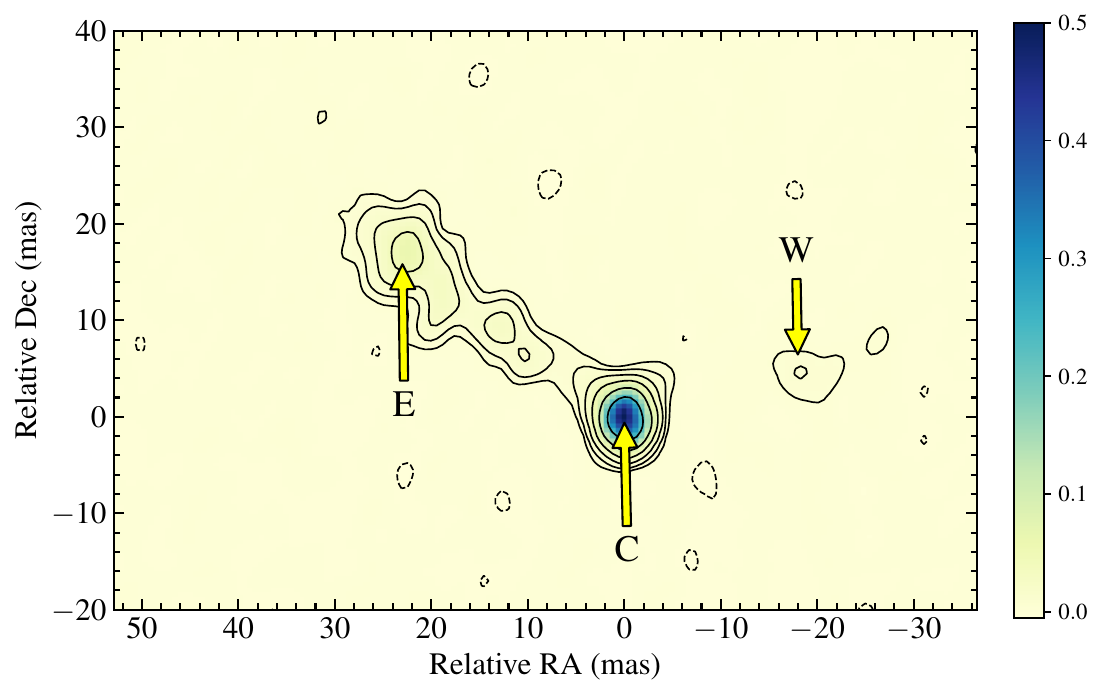}
    \includegraphics[scale=0.5]
    {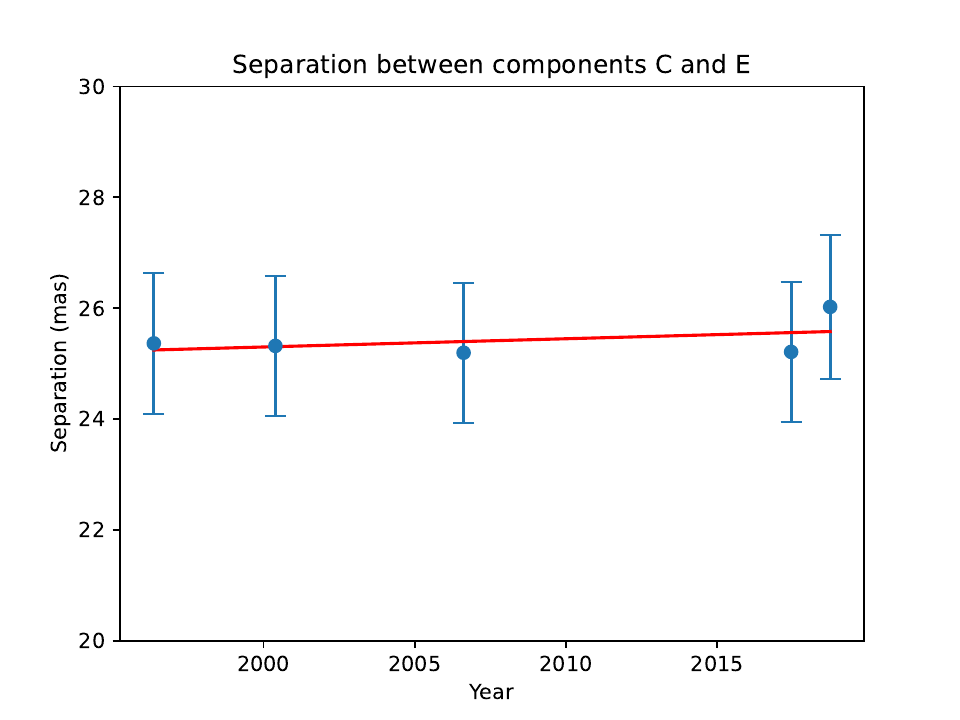}
   }
\hbox{
    \includegraphics[scale=0.48]
    {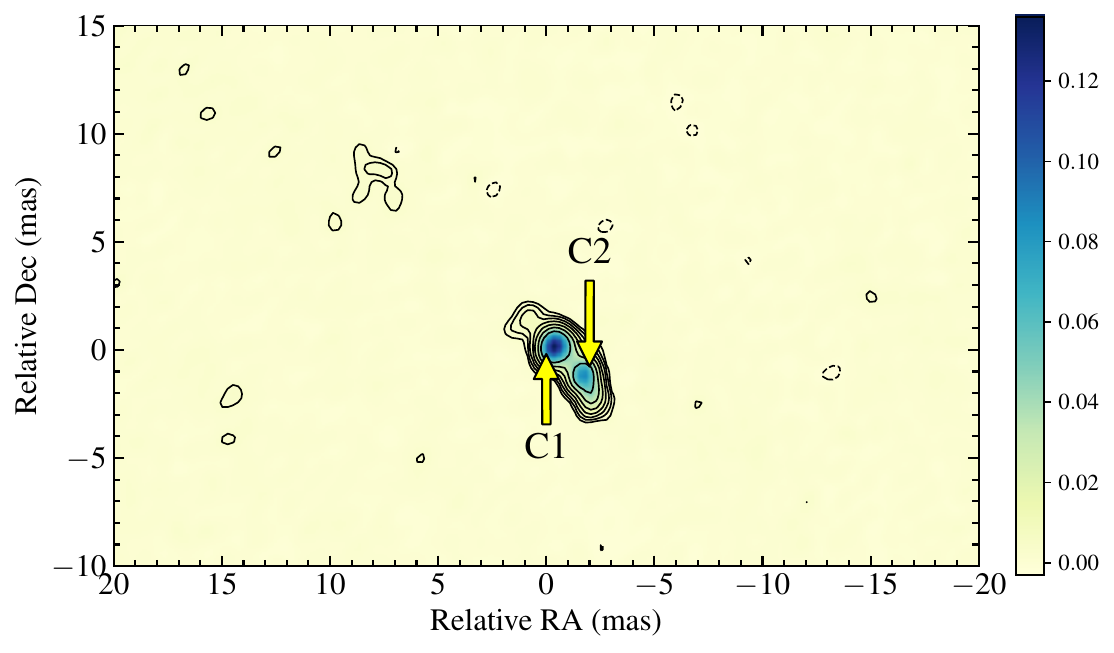}
    \includegraphics[scale=0.5]
    {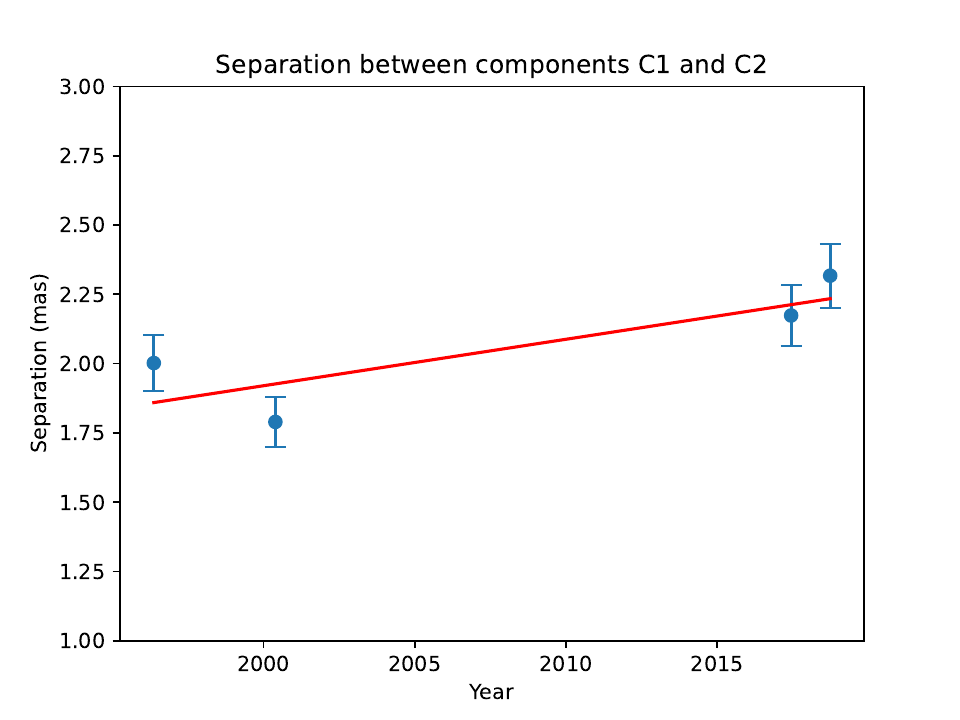}
    }
    \caption{The VLBA images of DA~362 at S-band (2.29 GHz, upper left) and X-band (8.65 GHz, lower left) from observations on 2000 May 22 (from astrogeo.org uploaded by Alexandr Pushkarev). The contour levels are 3$\sigma \times -1, 1, 2, 4, ...$ where $\sigma$ is 1.1 and 0.7 mJy/beam for the S- and X-band images respectively.  The linear least squares fit to the separations between components C and E (upper right) and C1 and C2 (lower right) are shown for the available data from astrogeo.org.}
    \label{fig:sublum}
\end{figure*}

The radio structure of DA 362 has been known to consist of a prominent jet towards the north-east, a weaker counter lobe on the western side, and a prominent central component \citep[cf.][and references therein]{2013MNRAS.433..147D}.
We utilized the five epochs of VLBI images covering 1996-2018 available in Astrogeo \citep[][]{2021AJ....161...14P} to determine the jet velocity. In the S-band and C-band observations, we determined the separation velocity between the central component C and the jet component E, which are well-imaged in all the maps at these frequencies (Figure~\ref{fig:sublum}). In the higher-frequency X-band, where the source has been observed in four epochs, the central component is resolved into a double, labeled C1 and C2 in Figure~\ref{fig:sublum}. It is unclear which of these two components may be hosting the nucleus of the galaxy. The apparent transverse velocity $\beta_{\rm app}=\mu d_\theta (1+z)/c$, where $\mu$ is the observed proper motion, $d_\theta$ the angular size distance, $z$ is the redshift and $c$ the speed of light. The least squares fit to the data show $\beta_{\rm app}$ between C and E, and C1 and C2 to be 0.21$\pm$0.30 and 0.24$\pm$0.16, respectively, consistent with a CSO and not a blazar. The separation between C and E is 102 pc, indicating a kinematic age of $\sim$1600 yr, assuming this velocity to be a constant. Extended emission beyond component E is seen, for example, in the lower-frequency L-band image \citep[][]{1995A&A...295...27D}, where they quote a total angular extent of 40 mas (161 pc) suggesting a kinematic age of $\sim$2500 yr for their outermost component. For DA 362, \citet[][]{2024ApJ...961..240K} reported the upper limit to the projected linear size to be 693 pc, considering the second lowest contours. On the other hand, we have reported the angular separation of the eastern component from the core by measuring it from the peak-to-peak positions of the core and the eastern hotspot.

\section{Discussion}\label{sec:dis}
\begin{figure*}
\hbox{
\includegraphics[scale=0.32]{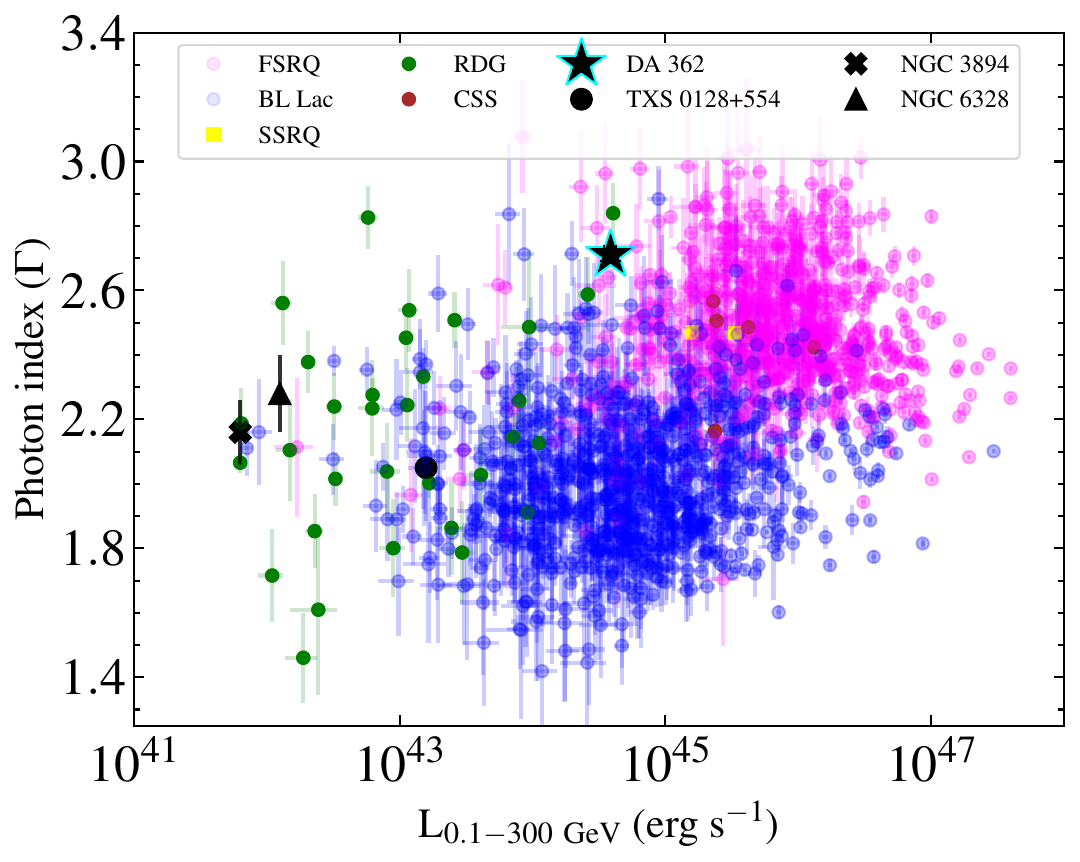}
\includegraphics[scale=0.32]{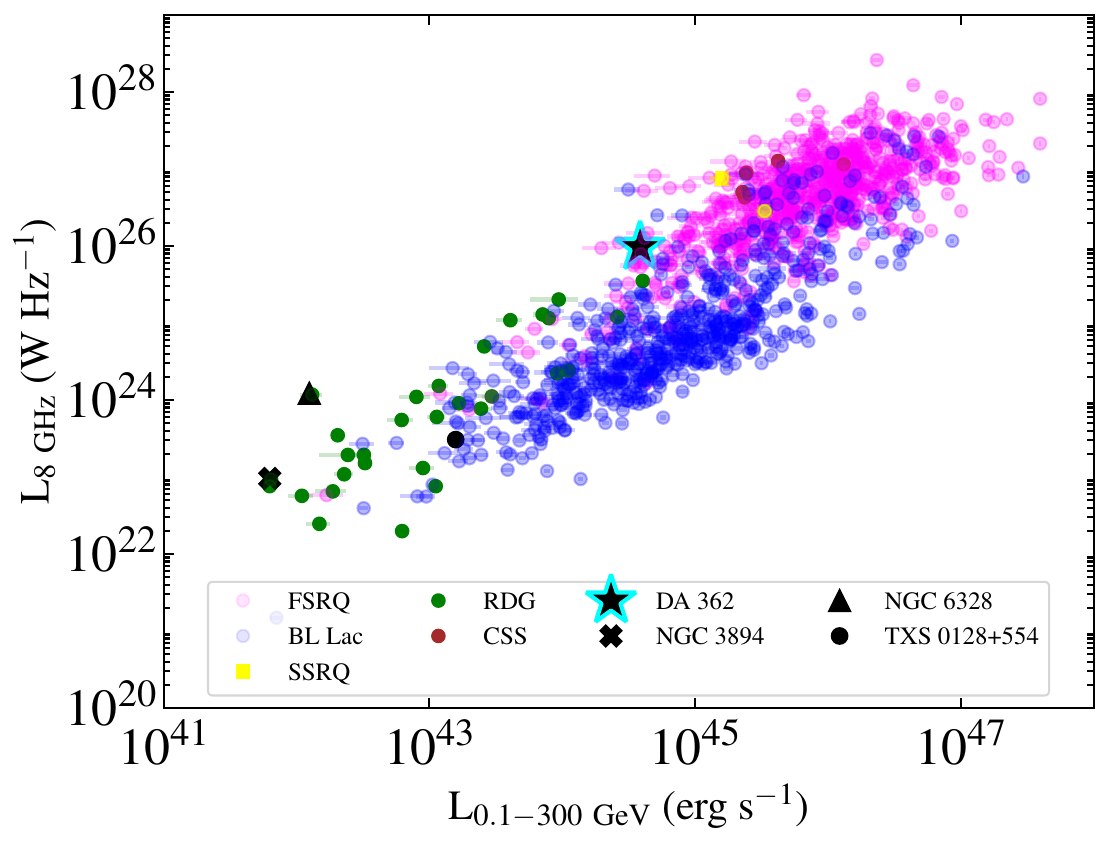}
\includegraphics[scale=0.33]{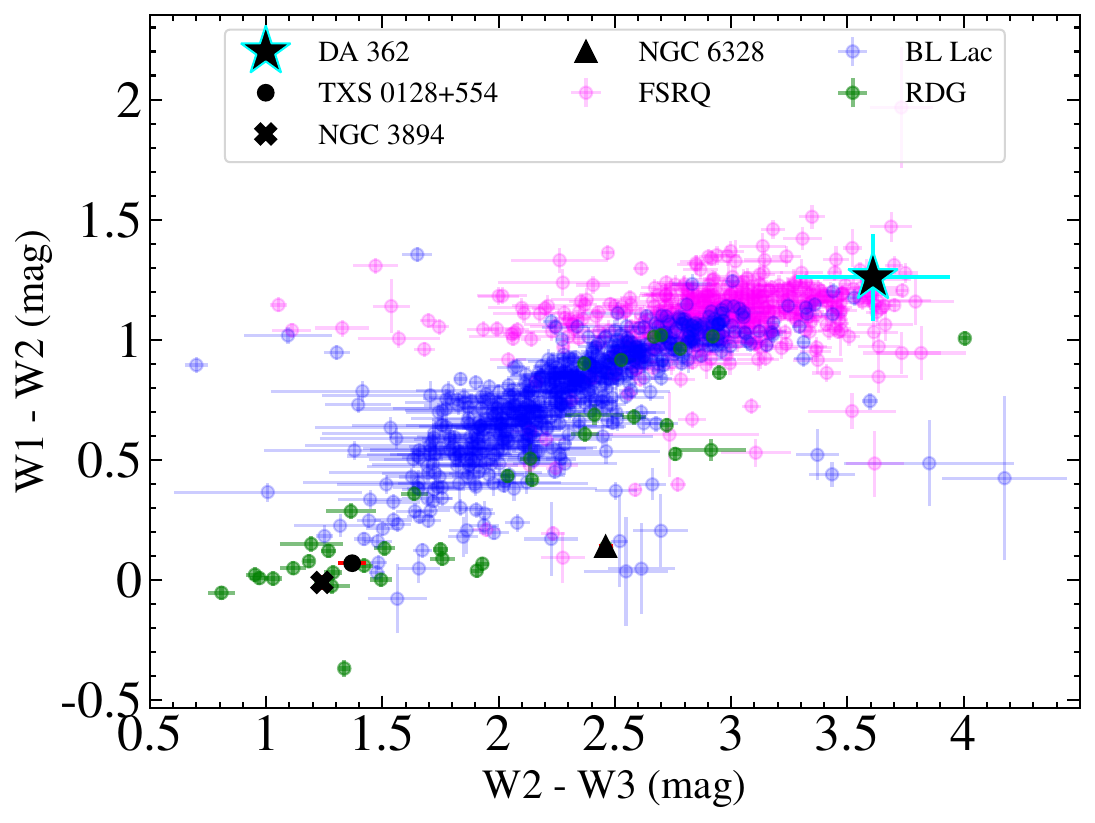}
}
\hbox{\hspace{1.5cm}
\includegraphics[scale=0.35]{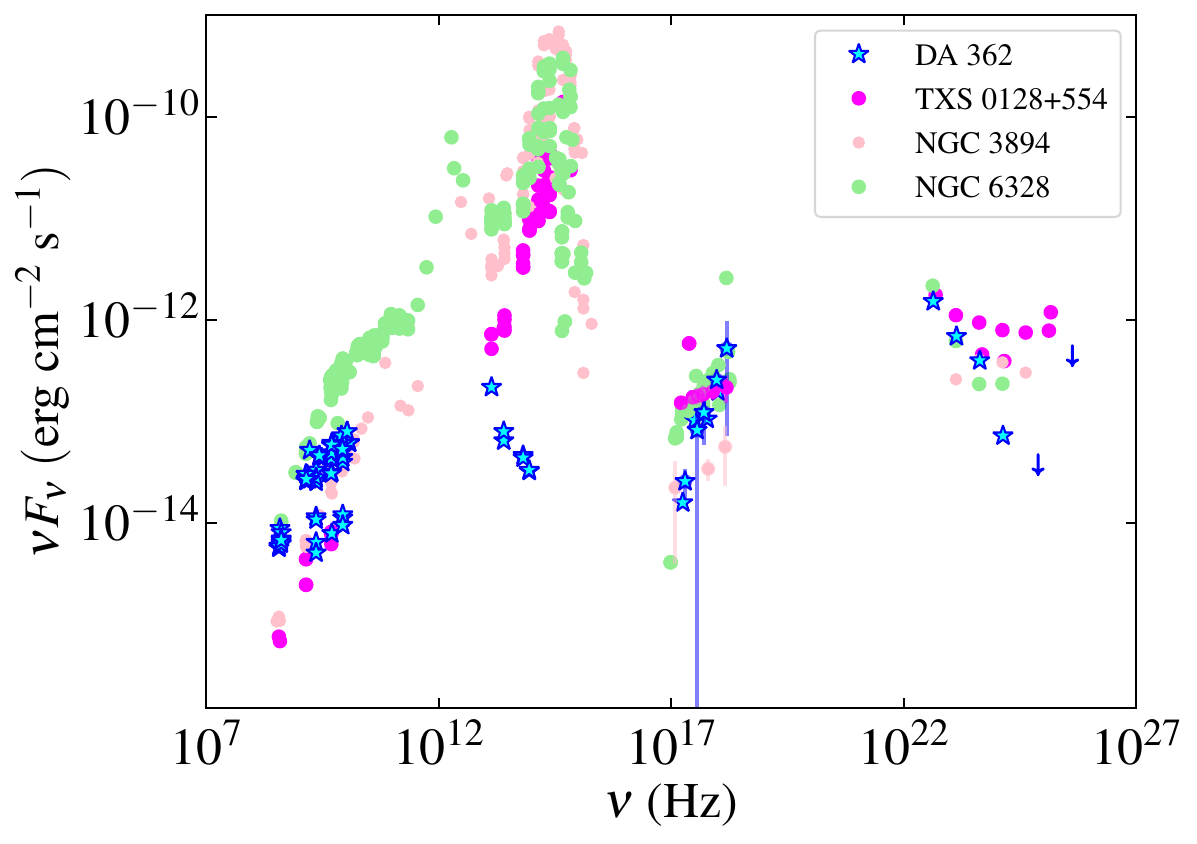}
\includegraphics[scale=0.35]{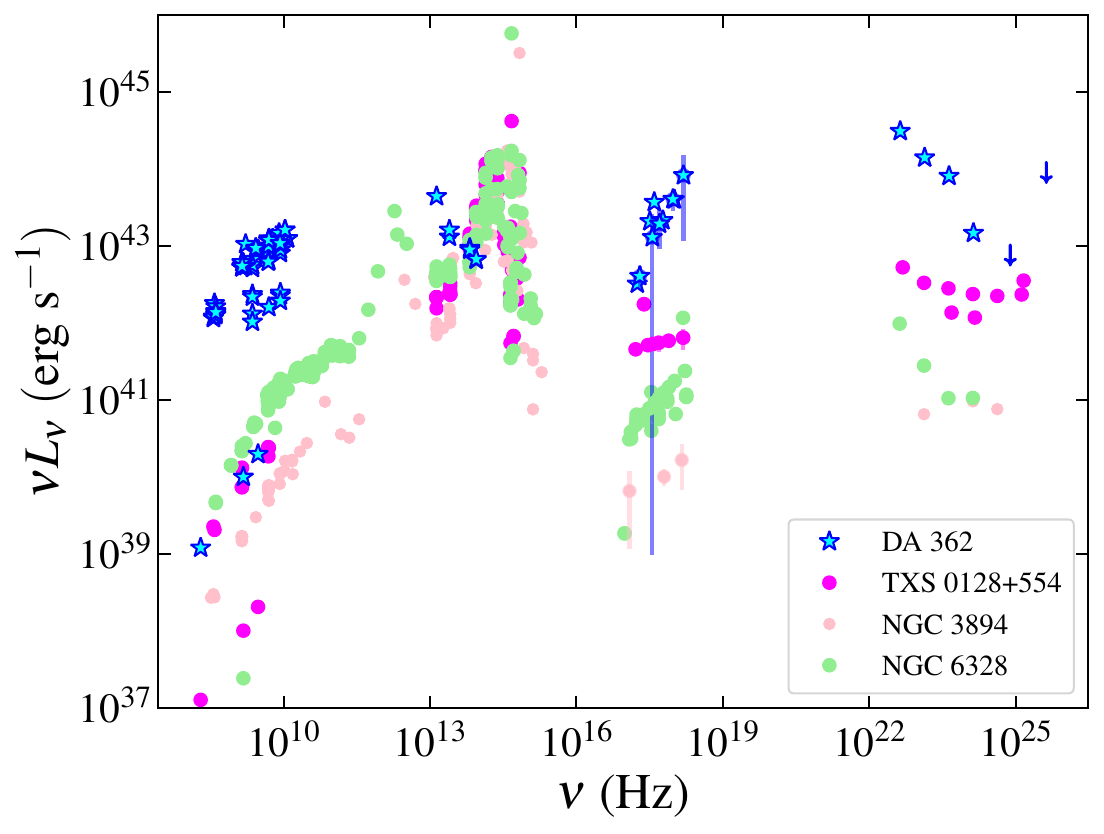}
}
\caption{The plots of the $\gamma$-ray luminosity vs. spectral index, and that of radio vs. \gm-ray luminosities are shown in the top left and middle panels. The top right panel shows the WISE color-color diagram. Blue and pink circles denote BL Lac objects and flat spectrum radio quasars, respectively. The yellow rectangle, green circles, and brown circles refer to steep spectrum radio quasars, radio galaxies and compact steep spectrum sources, respectively. These \gm-ray sources were selected from the 4FGL-DR4 catalog \citep[][]{2020ApJS..247...33A,ballet2023}. We also show \gm-ray detected CSOs, as labeled. The broadband SEDs of \gm-ray detected CSOs are plotted in the bottom panels.}\label{fig:l_ind}
\end{figure*}

The detection of small, parsec-scale, bi-polar radio emission and a subluminal motion, as revealed by the VLBI datasets, provide unambiguous confirmation that DA~362 is a bona fide CSO, thereby making it only the fourth \gm-ray detected object of this class. On comparing its \gm-ray spectral properties with the other three \gm-ray detected CSOs, we found that DA~362 is the brightest among them and exhibits a spectrum steeper than the other sources. In the \gm-ray luminosity versus photon index plane, DA~362 appears to lie in a region of high luminosity and soft spectrum (Figure~\ref{fig:l_ind}, top left panel). The \gm-ray luminosity of the source was calculated assuming $z=0.26$ \citep[][]{1992PASA...10..140W}. Given that the origin of the redshift is uncertain, a \gm-ray luminosity more than an order of magnitude larger than other CSOs should be treated with caution. We also compared the radio and \gm-ray luminosities of CSOs with other jetted AGNs in the top middle panel of Figure~\ref{fig:l_ind} adopting the 8 GHz VLBI flux densities reported in the Radio Fundamental Catalog (Petrov \& Kovalev, 2024, under review). In this diagram, DA~362 appears to follow the observed correlation between the radio and \gm-ray luminosities that have larger values compared to other \gm-ray detected CSOs.

\begin{figure}
    \includegraphics[width=\linewidth]{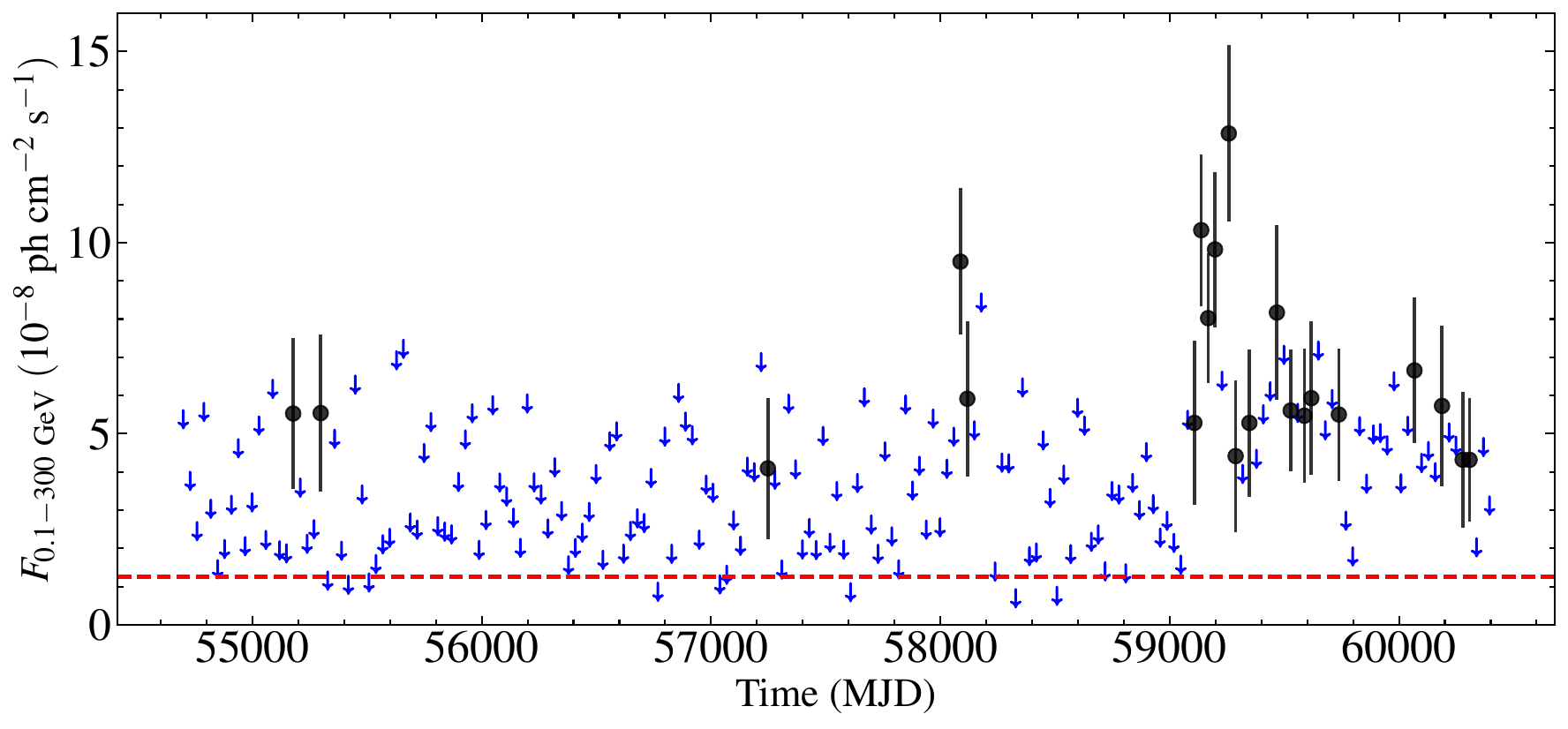}
    \caption{The monthly binned \gm-ray light curve of DA~362. The 95\% flux upper limits are shown with the downward arrows. The horizontal dashed line refers to the mission averaged \gm-ray flux of the source.}
    \label{fig:lc}
\end{figure}

\begin{figure}
    \includegraphics[width=\linewidth]{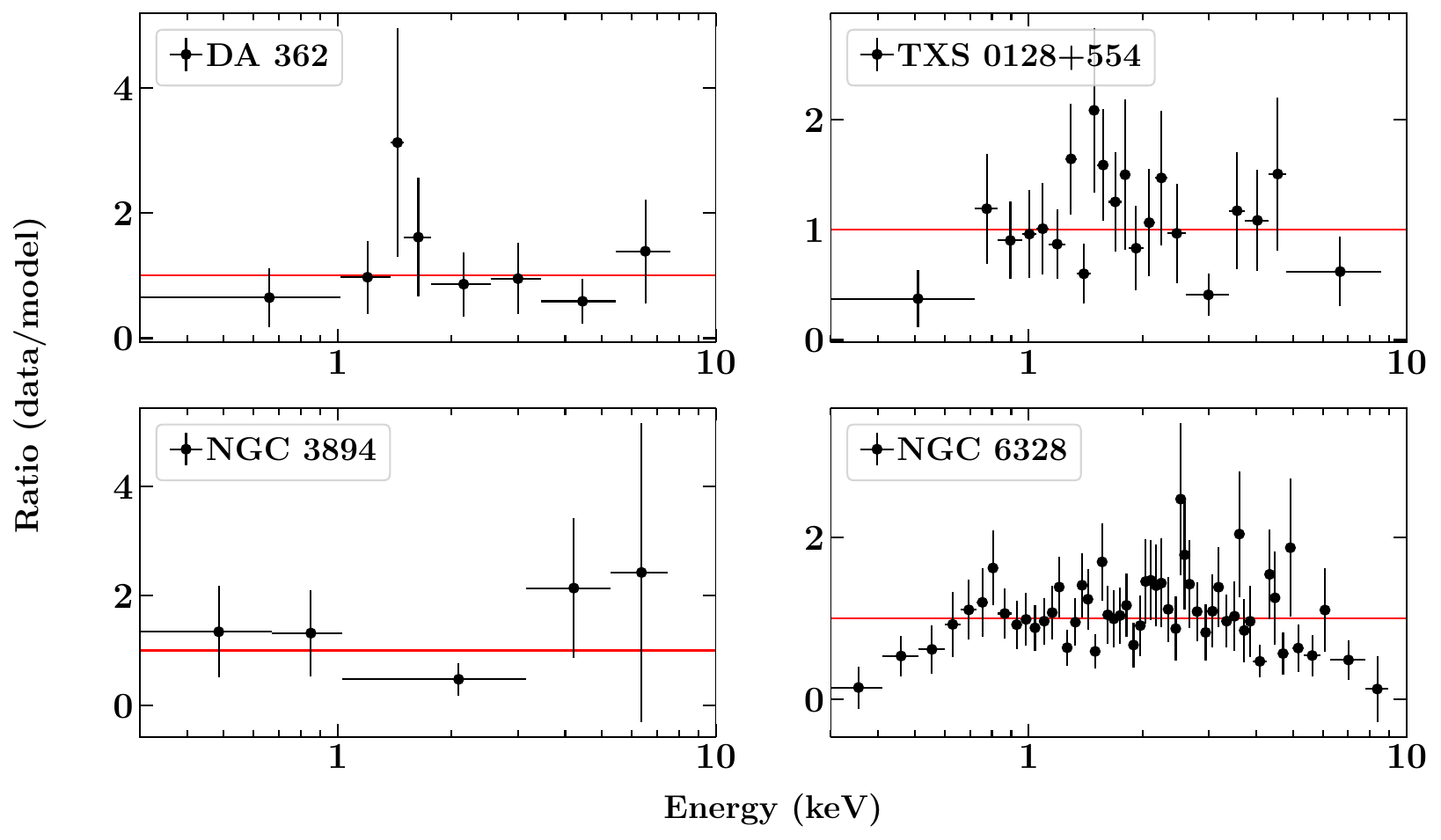}
    \caption{The residuals of the power law model fitting with Galactic absorption.}
    \label{fig:ratio}
\end{figure}

The \gm-ray emission detected from CSOs is not expected to exhibit significant variability if it originated from the radio lobes. Indeed, none of the three \gm-ray emitting CSOs have displayed any flux variability so far. However, a \gm-ray flaring activity of DA~362 was identified with the Fermi-LAT \citep[Figure~\ref{fig:lc}; see also][]{2021ApJS..256...13B}. This observation suggests the \gm-ray emission to be likely produced in the inner regions of the jet/core and not by the radio lobes. Similar results were found for another \gm-ray emitting CSO TXS 0128+554 \citep[][]{lister2020}. However, given the low photon statistics, it is not possible to quantify the flux variability timescales to derive meaningful constraints on the location of the emission region.

We considered the Swift-XRT data of other three \gm-ray emitting CSOs for a similar comparison and adopted the same data reduction and spectral fitting methodology described in Section~\ref{swift}.  In particular, TXS 0128+554 was serendipitously detected when Swift-XRT was observing GRB 190203a (target id: 88751). The Swift-XRT was pointed to NGC 3894 once (target id: 89108) and it was also serendipitously observed during Swift Gravitational Wave Galaxy Survey (target ids: 3107061, 3105839). On the other hand, NGC 6328 was observed 13 times (target id: 31815, 89109). We added individual observations to generate a combined X-ray spectrum for every CSO. The obtained net counts were 124, 12, and 413, for TXS 0128+554, NGC 3894, and NGC 6328, respectively. 

We fitted a model including a power law component and Galactic and intrinsic absorption. For TXS 128+554 and NGC 3894, the intrinsic neutral hydrogen column density could not be constrained, hence frozen to the values reported in recent works \citep[][]{lister2020,2021ApJ...922...84B}. For NGC 6328, on the other hand, the data quality was good enough to determine the intrinsic column density. It was estimated to be $1.39^{+0.80}_{-0.70}\times10^{21}$ cm$^{-2}$ which is consistent with that obtained by \citet[][]{2024A&A...684A..65B}. The obtained parameters are provided in Table~\ref{tab:tab0} and we show the residuals of the fit in Figure~\ref{fig:ratio}. The unfolded spectra, in $\nu F_{\nu}$ versus $\nu$ format, were extracted by rebinning them to have at least 5$\sigma$ detection in each bin or grouped in sets of 5 bins. The Swift-XRT spectral fitting results for NGC 6328 were published by \citet[][]{2023PASJ...75.1124M} and the spectral parameters estimated in this work are fully consistent with their published values. For all other three sources, the Swift-XRT data fitting results are reported in this work for the first time. We note that NGC 3894 and NGC 6328 have also been observed with XMM-Newton, Chandra, and/or NuSTAR, and more complex spectral modeling have been performed on them \citep[e.g.,][]{2021ApJ...922...84B,Sobolewska2022,2024A&A...684A..65B}. High-quality X-ray observations of DA 362, e.g., with XMM-Newton, will be needed to test physically motivated spectral models as done for other \gm-ray detected CSOs.

On comparing the Swift-XRT spectral parameters of DA~362 with those estimated for other \gm-ray detected CSOs, we found that it has absorption-corrected 0.3$-$10 keV flux brightness similar to other sources, though the uncertainties in the flux value are large, possibly due to low exposure. While other \gm-ray emitting CSOs have X-ray photon indices similar to that typically observed from \gm-ray emitting radio galaxies \citep[cf.][]{2023PASJ...75.1124M}, DA~362 exhibits an extremely hard X-ray spectrum. Usually, such a flat X-ray spectrum is observed from Compton-thick AGN due to severe absorption of the soft X-ray photons. Interestingly, the WISE observations indicate it to be bright in the MIR band, while the source is extremely faint in the optical-UV band, thus making it a very red object. Combining the MIR-to-UV photometric results with the X-ray spectral parameters, a picture emerges hinting at strong dust obscuration supporting the possible obscured nature of DA~362. Deeper X-ray observations will be crucial to test this scenario.

In the WISE color-color diagram of \gm-ray sources, DA~362 appears to lie in a region mainly populated by flat spectrum radio quasars (Figure~\ref{fig:l_ind}, top right panel). Therefore, its MIR emission could originate via synchrotron emission produced by compact radio lobes expanding with mildly relativistic velocities that may not be beamed. In contrast, other \gm-ray emitting CSOs are located in an area dominated by elliptical galaxies. A possible explanation could be that the MIR emission observed from these objects is primarily dominated by the thermal radiation from the host galaxy \citep[see also][]{2020ApJ...897..164K}. 
DA~362, on the other hand, lies in a region mainly occupied by quasars/Seyferts \citep[][]{2012MNRAS.426.3271M,2012ApJ...753...30S}, which also includes the 3 sources marked as Compton thick in \citet[][]{2020ApJ...897..164K}, as well as a few other CSOs.

The broadband spectral energy distribution (SED) of DA~362 is shown in the bottom left panel of Figure~\ref{fig:l_ind}. For comparison, we also plot the SEDs of the other three \gm-ray emitting CSOs that have also been studied in previous works \citep[][]{lister2020,2021ApJ...922...84B,Sobolewska2022,2024A&A...684A..65B}. At GHz frequencies, DA~362 has a brightness similar to TXS~0128+554 and NGC~3894, though fainter than NGC~6328. The primary difference can be seen in the MIR-to-UV energy range, where the observed emission is dominated by the host galaxy for other \gm-ray emitting CSOs, which is not the case for DA~362. Deep optical-MIR photometric and spectroscopic observations are needed to characterize its host galaxy properties and accretion activity. We discussed the comparison of the X- and \gm-ray properties earlier in this section, and the observed SEDs at these energies are consistent with the reported findings. We also show the multiwavelength SEDs of \gm-ray emitting CSOs in $\nu L_{\nu}$ versus $\nu$ plane in the bottom right panel of Figure~\ref{fig:l_ind}. Given the larger redshift of DA 362, it appeared more luminous compared to other \gm-ray detected CSOs at all wavelengths, except at optical frequencies where it is less luminous. However, since the redshift information of DA 362 is tentative, firm conclusions about its nature cannot be drawn.

\section{Summary}\label{sec:sum}
In this work, we have studied the multiwavelength properties of a CSO DA~362, which was recently found to be a \gm-ray emitter by the Fermi-LAT, thereby making it only the fourth \gm-ray detected object of this class of AGN. We summarize our key findings below:
\begin{enumerate}
    \item We confirm the association of the \gm-ray source 4FGL~J1416.0+3443 with DA~362 by analyzing $\sim$15.75 years of the Fermi-LAT data. The optimized \gm-ray position was consistent with the radio source within the estimated 95\% \gm-ray uncertainty region.
    \item The monthly binned \gm-ray light curve of DA~362 revealed a flaring activity during MJD 59075-59287. This is the first detection of a \gm-ray flare from a CSO which was also reported by \citet[][]{2021ApJS..256...13B}. This peculiar flaring activity indicates that the \gm-ray emission to originate from the core/jet rather than from the radio lobes.
    \item The source exhibits an extremely hard X-ray spectrum (0.3$-$10 keV photon index = 0.79$^{+0.52}_{-0.46}$) as found with the analysis of the low-exposure Swift-XRT data. However, a strong claim cannot be made due to large uncertainties.
    \item DA~362 is bright in MIR but extremely faint in the optical band, thus suggesting possible dust obscuration. By also considering the observed X-ray spectral shape, these results indicate a possible X-ray obscured nature of the source.
    \item We used the calibrated VLBI images from the Astrogeo website and estimated the jet separation velocity to be $v_{\rm app}\sim 0.2c$. This detection of a subluminal motion further supports the CSO nature of DA~362.
    \item The available observations have provided tantalizing clues about the enigmatic behavior of this \gm-ray emitting CSO. Deeper observations with sensitive observing facilities will be needed to explore the broadband physical properties of DA~362 and probe the origin of \gm-ray emission.
\end{enumerate}

\begin{acknowledgements}
We thank the journal referee for constructive criticism. Thanks to Swift Satellite's principal investigator, Brad Cenko, for approving the observation request. Part of this work is based on archival data, software, or online services provided by the Space Science Data Center - ASI. This research has made use of NASA’s Astrophysics Data System Bibliographic Services. This research has made use of data obtained through the High Energy Astrophysics Science Archive Research Center Online Service, provided by the NASA/Goddard Space Flight Center. 
\end{acknowledgements}
\facilities{Swift, Fermi-LAT}

\software{XSPEC \citep[v 12.10.1;][]{1996ASPC..101...17A}, Swift-XRT data product generator \citep[][]{2009MNRAS.397.1177E}, fermiPy \citep{2017arXiv170709551W}}

\bibliography{reference}{}

\begin{thebibliography}{}
\expandafter\ifx\csname natexlab\endcsname\relax\def\natexlab#1{#1}\fi
\providecommand{\url}[1]{\href{#1}{#1}}
\providecommand{\dodoi}[1]{doi:~\href{http://doi.org/#1}{\nolinkurl{#1}}}
\providecommand{\doeprint}[1]{\href{http://ascl.net/#1}{\nolinkurl{http://ascl.net/#1}}}
\providecommand{\doarXiv}[1]{\href{https://arxiv.org/abs/#1}{\nolinkurl{https://arxiv.org/abs/#1}}}

\bibitem[{{Abdollahi} {et~al.}(2020){Abdollahi}, {Acero}, {Ackermann},
  {Ajello}, {Atwood}, {Axelsson}, {Baldini}, {Ballet}, {Barbiellini},
  {Bastieri}, {Becerra Gonzalez}, {Bellazzini}, {Berretta}, {Bissaldi},
  {Blandford}, {Bloom}, {Bonino}, {Bottacini}, {Brandt}, {Bregeon}, {Bruel},
  {Buehler}, {Burnett}, {Buson}, {Cameron}, {Caputo}, {Caraveo}, {Casandjian},
  {Castro}, {Cavazzuti}, {Charles}, {Chaty}, {Chen}, {Cheung}, {Chiaro},
  {Ciprini}, {Cohen-Tanugi}, {Cominsky}, {Coronado-Bl{\'a}zquez}, {Costantin},
  {Cuoco}, {Cutini}, {D'Ammando}, {DeKlotz}, {de la Torre Luque}, {de Palma},
  {Desai}, {Digel}, {Di Lalla}, {Di Mauro}, {Di Venere}, {Dom{\'\i}nguez},
  {Dumora}, {Fana Dirirsa}, {Fegan}, {Ferrara}, {Franckowiak}, {Fukazawa},
  {Funk}, {Fusco}, {Gargano}, {Gasparrini}, {Giglietto}, {Giommi}, {Giordano},
  {Giroletti}, {Glanzman}, {Green}, {Grenier}, {Griffin}, {Grondin}, {Grove},
  {Guiriec}, {Harding}, {Hayashi}, {Hays}, {Hewitt}, {Horan},
  {J{\'o}hannesson}, {Johnson}, {Kamae}, {Kerr}, {Kocevski}, {Kovac'evic'},
  {Kuss}, {Landriu}, {Larsson}, {Latronico}, {Lemoine-Goumard}, {Li},
  {Liodakis}, {Longo}, {Loparco}, {Lott}, {Lovellette}, {Lubrano}, {Madejski},
  {Maldera}, {Malyshev}, {Manfreda}, {Marchesini}, {Marcotulli},
  {Mart{\'\i}-Devesa}, {Martin}, {Massaro}, {Mazziotta}, {McEnery}, {Mereu},
  {Meyer}, {Michelson}, {Mirabal}, {Mizuno}, {Monzani}, {Morselli},
  {Moskalenko}, {Negro}, {Nuss}, {Ojha}, {Omodei}, {Orienti}, {Orlando},
  {Ormes}, {Palatiello}, {Paliya}, {Paneque}, {Pei}, {Pe{\~n}a-Herazo},
  {Perkins}, {Persic}, {Pesce-Rollins}, {Petrosian}, {Petrov}, {Piron}, {Poon},
  {Porter}, {Principe}, {Rain{\`o}}, {Rando}, {Razzano}, {Razzaque}, {Reimer},
  {Reimer}, {Remy}, {Reposeur}, {Romani}, {Saz Parkinson}, {Schinzel},
  {Serini}, {Sgr{\`o}}, {Siskind}, {Smith}, {Spandre}, {Spinelli}, {Strong},
  {Suson}, {Tajima}, {Takahashi}, {Tak}, {Thayer}, {Thompson}, {Tibaldo},
  {Torres}, {Torresi}, {Valverde}, {Van Klaveren}, {van Zyl}, {Wood},
  {Yassine}, \& {Zaharijas}}]{2020ApJS..247...33A}
{Abdollahi}, S., {Acero}, F., {Ackermann}, M., {et~al.} 2020, The Astrophysical
  Journal Supplement Series, 247, 33, \dodoi{10.3847/1538-4365/ab6bcb}

\bibitem[{{Abdollahi} {et~al.}(2023){Abdollahi}, {Ajello}, {Baldini}, {Ballet},
  {Bastieri}, {Becerra Gonzalez}, {Bellazzini}, {Berretta}, {Bissaldi},
  {Bonino}, {Brill}, {Bruel}, {Burns}, {Buson}, {Cameron}, {Caputo}, {Caraveo},
  {Cibrario}, {Ciprini}, {Cristarella Orestano}, {Crnogorcevic}, {Cutini},
  {D'Ammando}, {De Gaetano}, {Digel}, {Di Lalla}, {Di Venere},
  {Dom{\'\i}nguez}, {Ramazani}, {Fegan}, {Ferrara}, {Fiori}, {Fleischhack},
  {Franckowiak}, {Fukazawa}, {Fusco}, {Gammaldi}, {Gargano}, {Garrappa},
  {Gasbarra}, {Gasparrini}, {Giglietto}, {Giordano}, {Giroletti}, {Green},
  {Grenier}, {Guiriec}, {Gustafsson}, {Hays}, {Horan}, {Hou},
  {J{\'o}hannesson}, {Kerr}, {Kocevski}, {Kuss}, {Latronico}, {Li}, {Liodakis},
  {Longo}, {Loparco}, {Lorusso}, {Lott}, {Lovellette}, {Lubrano}, {Maldera},
  {Manfreda}, {Mart{\'\i}-Devesa}, {Mazziotta}, {Mereu}, {Meyer}, {Michelson},
  {Mizuno}, {Monzani}, {Morselli}, {Moskalenko}, {Negro}, {Omodei}, {Orlando},
  {Ormes}, {Paneque}, {Panzarini}, {Perkins}, {Persic}, {Pesce-Rollins},
  {Pillera}, {Porter}, {Principe}, {Racusin}, {Rain{\`o}}, {Rando}, {Rani},
  {Razzano}, {Razzaque}, {Reimer}, {Reimer}, {S{\'a}nchez-Conde}, {Parkinson},
  {Scargle}, {Scotton}, {Serini}, {Sgr{\`o}}, {Siskind}, {Spandre}, {Spinelli},
  {Suson}, {Tajima}, {Thompson}, {Torres}, {Valverde}, {Venters}, {Wadiasingh},
  {Wagner}, \& {Wood}}]{2023ApJS..265...31A}
{Abdollahi}, S., {Ajello}, M., {Baldini}, L., {et~al.} 2023, \apjs, 265, 31,
  \dodoi{10.3847/1538-4365/acbb6a}

\bibitem[{{Ahumada} {et~al.}(2020){Ahumada}, {Allende Prieto}, {Almeida},
  {Anders}, {Anderson}, {Andrews}, {Anguiano}, {Arcodia}, {Armengaud},
  {Aubert}, {Avila}, {Avila-Reese}, {Badenes}, {Balland}, {Barger},
  {Barrera-Ballesteros}, {Basu}, {Bautista}, {Beaton}, {Beers}, {Benavides},
  {Bender}, {Bernardi}, {Bershady}, {Beutler}, {Bidin}, {Bird}, {Bizyaev},
  {Blanc}, {Blanton}, {Boquien}, {Borissova}, {Bovy}, {Brandt}, {Brinkmann},
  {Brownstein}, {Bundy}, {Bureau}, {Burgasser}, {Burtin}, {Cano-D{\'\i}az},
  {Capasso}, {Cappellari}, {Carrera}, {Chabanier}, {Chaplin}, {Chapman},
  {Cherinka}, {Chiappini}, {Doohyun Choi}, {Chojnowski}, {Chung}, {Clerc},
  {Coffey}, {Comerford}, {Comparat}, {da Costa}, {Cousinou}, {Covey}, {Crane},
  {Cunha}, {Ilha}, {Dai}, {Damsted}, {Darling}, {Davidson}, {Davies}, {Dawson},
  {De}, {de la Macorra}, {De Lee}, {Queiroz}, {Deconto Machado}, {de la Torre},
  {Dell'Agli}, {du Mas des Bourboux}, {Diamond-Stanic}, {Dillon}, {Donor},
  {Drory}, {Duckworth}, {Dwelly}, {Ebelke}, {Eftekharzadeh}, {Davis Eigenbrot},
  {Elsworth}, {Eracleous}, {Erfanianfar}, {Escoffier}, {Fan}, {Farr},
  {Fern{\'a}ndez-Trincado}, {Feuillet}, {Finoguenov}, {Fofie},
  {Fraser-McKelvie}, {Frinchaboy}, {Fromenteau}, {Fu}, {Galbany}, {Garcia},
  {Garc{\'\i}a-Hern{\'a}ndez}, {Garma Oehmichen}, {Ge}, {Geimba Maia},
  {Geisler}, {Gelfand}, {Goddy}, {Gonzalez-Perez}, {Grabowski}, {Green},
  {Grier}, {Guo}, {Guy}, {Harding}, {Hasselquist}, {Hawken}, {Hayes}, {Hearty},
  {Hekker}, {Hogg}, {Holtzman}, {Horta}, {Hou}, {Hsieh}, {Huber}, {Hunt}, {Ider
  Chitham}, {Imig}, {Jaber}, {Jimenez Angel}, {Johnson}, {Jones},
  {J{\"o}nsson}, {Jullo}, {Kim}, {Kinemuchi}, {Kirkpatrick}, {Kite}, {Klaene},
  {Kneib}, {Kollmeier}, {Kong}, {Kounkel}, {Krishnarao}, {Lacerna}, {Lan},
  {Lane}, {Law}, {Le Goff}, {Leung}, {Lewis}, {Li}, {Lian}, {Lin}, {Long},
  {Longa-Pe{\~n}a}, {Lundgren}, {Lyke}, {Mackereth}, {MacLeod}, {Majewski},
  {Manchado}, {Maraston}, {Martini}, {Masseron}, {Masters}, {Mathur},
  {McDermid}, {Merloni}, {Merrifield}, {M{\'e}sz{\'a}ros}, {Miglio}, {Minniti},
  {Minsley}, {Miyaji}, {Mohammad}, {Mosser}, {Mueller}, {Muna},
  {Mu{\~n}oz-Guti{\'e}rrez}, {Myers}, {Nadathur}, {Nair}, {Nandra}, {Correa do
  Nascimento}, {Nevin}, {Newman}, {Nidever}, {Nitschelm}, {Noterdaeme},
  {O'Connell}, {Olmstead}, {Oravetz}, {Oravetz}, {Osorio}, {Pace}, {Padilla},
  {Palanque-Delabrouille}, {Palicio}, {Pan}, {Pan}, {Parker}, {Paviot},
  {Peirani}, {Ram{\'r}ez}, {Penny}, {Percival}, {Perez-Fournon},
  {P{\'e}rez-R{\`a}fols}, {Petitjean}, {Pieri}, {Pinsonneault}, {Poovelil},
  {Povick}, {Prakash}, {Price-Whelan}, {Raddick}, {Raichoor}, {Ray}, {Rembold},
  {Rezaie}, {Riffel}, {Riffel}, {Rix}, {Robin}, {Roman-Lopes},
  {Rom{\'a}n-Z{\'u}{\~n}iga}, {Rose}, {Ross}, {Rossi}, {Rowlands}, {Rubin},
  {Salvato}, {S{\'a}nchez}, {S{\'a}nchez-Menguiano}, {S{\'a}nchez-Gallego},
  {Sayres}, {Schaefer}, {Schiavon}, {Schimoia}, {Schlafly}, {Schlegel},
  {Schneider}, {Schultheis}, {Schwope}, {Seo}, {Serenelli}, {Shafieloo},
  {Shamsi}, {Shao}, {Shen}, {Shetrone}, {Shirley}, {Silva Aguirre}, {Simon},
  {Skrutskie}, {Slosar}, {Smethurst}, {Sobeck}, {Sodi}, {Souto}, {Stark},
  {Stassun}, {Steinmetz}, {Stello}, {Stermer}, {Storchi-Bergmann},
  {Streblyanska}, {Stringfellow}, {Stutz}, {Su{\'a}rez}, {Sun},
  {Taghizadeh-Popp}, {Talbot}, {Tayar}, {Thakar}, {Theriault}, {Thomas},
  {Thomas}, {Tinker}, {Tojeiro}, {Toledo}, {Tremonti}, {Troup}, {Tuttle},
  {Unda-Sanzana}, {Valentini}, {Vargas-Gonz{\'a}lez}, {Vargas-Maga{\~n}a},
  {V{\'a}zquez-Mata}, {Vivek}, {Wake}, {Wang}, {Weaver}, {Weijmans}, {Wild},
  {Wilson}, {Wilson}, {Wolthuis}, {Wood-Vasey}, {Yan}, {Yang}, {Y{\`e}che},
  {Zamora}, {Zarrouk}, {Zasowski}, {Zhang}, {Zhao}, {Zhao}, {Zheng}, {Zheng},
  {Zhu}, \& {Zou}}]{2020ApJS..249....3A}
{Ahumada}, R., {Allende Prieto}, C., {Almeida}, A., {et~al.} 2020, \apjs, 249,
  3, \dodoi{10.3847/1538-4365/ab929e}

\bibitem[{{Arnaud}(1996)}]{1996ASPC..101...17A}
{Arnaud}, K.~A. 1996, in Astronomical Society of the Pacific Conference Series,
  Vol. 101, Astronomical Data Analysis Software and Systems V, ed. G.~H.
  {Jacoby} \& J.~{Barnes}, 17

\bibitem[{{Balasubramaniam} {et~al.}(2021){Balasubramaniam}, {Stawarz},
  {Cheung}, {Sobolewska}, {Marchenko}, {Thimmappa}, {Kr{\'o}l}, {Migliori}, \&
  {Siemiginowska}}]{2021ApJ...922...84B}
{Balasubramaniam}, K., {Stawarz}, {\L}., {Cheung}, C.~C., {et~al.} 2021, \apj,
  922, 84, \dodoi{10.3847/1538-4357/ac1ff5}

\bibitem[{{Baldini} {et~al.}(2021){Baldini}, {Ballet}, {Bastieri}, {Becerra
  Gonzalez}, {Bellazzini}, {Berretta}, {Bissaldi}, {Blandford}, {Bloom},
  {Bonino}, {Bottacini}, {Bruel}, {Buson}, {Cameron}, {Caraveo}, {Cavazzuti},
  {Chen}, {Chiaro}, {Ciangottini}, {Cibario}, {Ciprini}, {Cristarella
  Orestano}, {Crnogorcevic}, {Cutini}, {D'Ammando}, {de la Torre Luque}, {de
  Palma}, {Digel}, {Di Lalla}, {Dirirsa}, {Di Venere}, {Dom{\'\i}nguez},
  {Fiori}, {Fleischhack}, {Franckowiak}, {Fukazawa}, {Funk}, {Fusco},
  {Gargano}, {Gasparrini}, {Germani}, {Giglietto}, {Giordano}, {Giroletti},
  {Green}, {Grenier}, {Griffin}, {Guiriec}, {Gustafsson}, {Hewitt}, {Horan},
  {Imazawa}, {J{\'o}hannesson}, {Kerr}, {Kocevski}, {Kuss}, {Larsson},
  {Latronico}, {Li}, {Liodakis}, {Longo}, {Loparco}, {Lovellette}, {Lubrano},
  {Maldera}, {Manfreda}, {Mart{\'\i}-Devesa}, {Matake}, {Mazziotta}, {Mereu},
  {Meyer}, {Mirabal}, {Mitthumsiri}, {Mizuno}, {Monzani}, {Morselli},
  {Moskalenko}, {Nagasawa}, {Negro}, {Ojha}, {Orienti}, {Orlando},
  {Palatiello}, {Paliya}, {Paneque}, {Pei}, {Persic}, {Pesce-Rollins},
  {Petrosian}, {Poon}, {Porter}, {Principe}, {Racusin}, {Rain{\`o}}, {Rando},
  {Rani}, {Razzano}, {Razzaque}, {Reimer}, {Reimer}, {Saz Parkinson},
  {Scotton}, {Serini}, {Sgr{\`o}}, {Siskind}, {Spandre}, {Spinelli}, {Suson},
  {Tajima}, {Tak}, {Torres}, {Tosti}, {Troja}, {Wood}, {Yassine}, {Zaharijas},
  \& {Fermi-LAT Collaboration}}]{2021ApJS..256...13B}
{Baldini}, L., {Ballet}, J., {Bastieri}, D., {et~al.} 2021, \apjs, 256, 13,
  \dodoi{10.3847/1538-4365/ac072a}

\bibitem[{{Ballet} {et~al.}(2023){Ballet}, {Bruel}, {Burnett}, {Lott}, \& {The
  Fermi-LAT collaboration}}]{ballet2023}
{Ballet}, J., {Bruel}, P., {Burnett}, T.~H., {Lott}, B., \& {The Fermi-LAT
  collaboration}. 2023, arXiv e-prints, arXiv:2307.12546,
  \dodoi{10.48550/arXiv.2307.12546}

\bibitem[{{Bronzini} {et~al.}(2024){Bronzini}, {Migliori}, {Vignali},
  {Sobolewska}, {Stawarz}, {Siemiginowska}, {Orienti}, {D'Ammando},
  {Giroletti}, {Principe}, \& {Balasubramaniam}}]{2024A&A...684A..65B}
{Bronzini}, E., {Migliori}, G., {Vignali}, C., {et~al.} 2024, \aap, 684, A65,
  \dodoi{10.1051/0004-6361/202348208}

\bibitem[{{Chambers} {et~al.}(2016){Chambers}, {Magnier}, {Metcalfe},
  {Flewelling}, {Huber}, {Waters}, {Denneau}, {Draper}, {Farrow}, {Finkbeiner},
  {Holmberg}, {Koppenhoefer}, {Price}, {Rest}, {Saglia}, {Schlafly}, {Smartt},
  {Sweeney}, {Wainscoat}, {Burgett}, {Chastel}, {Grav}, {Heasley}, {Hodapp},
  {Jedicke}, {Kaiser}, {Kudritzki}, {Luppino}, {Lupton}, {Monet}, {Morgan},
  {Onaka}, {Shiao}, {Stubbs}, {Tonry}, {White}, {Ba{\~n}ados}, {Bell},
  {Bender}, {Bernard}, {Boegner}, {Boffi}, {Botticella}, {Calamida},
  {Casertano}, {Chen}, {Chen}, {Cole}, {Deacon}, {Frenk}, {Fitzsimmons},
  {Gezari}, {Gibbs}, {Goessl}, {Goggia}, {Gourgue}, {Goldman}, {Grant},
  {Grebel}, {Hambly}, {Hasinger}, {Heavens}, {Heckman}, {Henderson}, {Henning},
  {Holman}, {Hopp}, {Ip}, {Isani}, {Jackson}, {Keyes}, {Koekemoer}, {Kotak},
  {Le}, {Liska}, {Long}, {Lucey}, {Liu}, {Martin}, {Masci}, {McLean}, {Mindel},
  {Misra}, {Morganson}, {Murphy}, {Obaika}, {Narayan}, {Nieto-Santisteban},
  {Norberg}, {Peacock}, {Pier}, {Postman}, {Primak}, {Rae}, {Rai}, {Riess},
  {Riffeser}, {Rix}, {R{\"o}ser}, {Russel}, {Rutz}, {Schilbach}, {Schultz},
  {Scolnic}, {Strolger}, {Szalay}, {Seitz}, {Small}, {Smith}, {Soderblom},
  {Taylor}, {Thomson}, {Taylor}, {Thakar}, {Thiel}, {Thilker}, {Unger},
  {Urata}, {Valenti}, {Wagner}, {Walder}, {Walter}, {Watters}, {Werner},
  {Wood-Vasey}, \& {Wyse}}]{2016arXiv161205560C}
{Chambers}, K.~C., {Magnier}, E.~A., {Metcalfe}, N., {et~al.} 2016, arXiv
  e-prints, arXiv:1612.05560, \dodoi{10.48550/arXiv.1612.05560}

\bibitem[{{Condon} {et~al.}(1998){Condon}, {Cotton}, {Greisen}, {Yin},
  {Perley}, {Taylor}, \& {Broderick}}]{1998AJ....115.1693C}
{Condon}, J.~J., {Cotton}, W.~D., {Greisen}, E.~W., {et~al.} 1998, \aj, 115,
  1693, \dodoi{10.1086/300337}

\bibitem[{{Dallacasa} {et~al.}(1995){Dallacasa}, {Fanti}, {Fanti}, {Schilizzi},
  \& {Spencer}}]{1995A&A...295...27D}
{Dallacasa}, D., {Fanti}, C., {Fanti}, R., {Schilizzi}, R.~T., \& {Spencer},
  R.~E. 1995, \aap, 295, 27

\bibitem[{{Dallacasa} {et~al.}(2013){Dallacasa}, {Orienti}, {Fanti}, {Fanti},
  \& {Stanghellini}}]{2013MNRAS.433..147D}
{Dallacasa}, D., {Orienti}, M., {Fanti}, C., {Fanti}, R., \& {Stanghellini}, C.
  2013, \mnras, 433, 147, \dodoi{10.1093/mnras/stt710}

\bibitem[{{Duchesne} {et~al.}(2024){Duchesne}, {Grundy}, {Heald}, {Lenc},
  {Leung}, {McConnell}, {Murphy}, {Pritchard}, {Rose}, {Thomson}, {Wang},
  {Wang}, \& {Whiting}}]{2024PASA...41....3D}
{Duchesne}, S.~W., {Grundy}, J.~A., {Heald}, G.~H., {et~al.} 2024, \pasa, 41,
  e003, \dodoi{10.1017/pasa.2023.60}

\bibitem[{{Epchtein} {et~al.}(1994){Epchtein}, {de Batz}, {Copet}, {Fouque},
  {Lacombe}, {Le Bertre}, {Mamon}, {Rouan}, {Tiphene}, {Burton}, {Deul},
  {Habing}, {Boersenberger}, {Dennefeld}, {Omont}, {Renault},
  {Rocca-Volmerange}, {Kimeswenger}, {Appenzeller}, {Bender}, {Forveille},
  {Garzon}, {Hron}, {Persi}, {Ferrari-Toniolo}, \&
  {Vauglin}}]{1994Ap&SS.217....3E}
{Epchtein}, N., {de Batz}, B., {Copet}, E., {et~al.} 1994, \apss, 217, 3,
  \dodoi{10.1007/BF00990013}

\bibitem[{{Evans} {et~al.}(2009){Evans}, {Beardmore}, {Page}, {Osborne},
  {O'Brien}, {Willingale}, {Starling}, {Burrows}, {Godet}, {Vetere}, {Racusin},
  {Goad}, {Wiersema}, {Angelini}, {Capalbi}, {Chincarini}, {Gehrels}, {Kennea},
  {Margutti}, {Morris}, {Mountford}, {Pagani}, {Perri}, {Romano}, \&
  {Tanvir}}]{2009MNRAS.397.1177E}
{Evans}, P.~A., {Beardmore}, A.~P., {Page}, K.~L., {et~al.} 2009, \mnras, 397,
  1177, \dodoi{10.1111/j.1365-2966.2009.14913.x}

\bibitem[{{Georgantopoulos} {et~al.}(2007){Georgantopoulos}, {Georgakakis}, \&
  {Akylas}}]{2007A&A...466..823G}
{Georgantopoulos}, I., {Georgakakis}, A., \& {Akylas}, A. 2007, \aap, 466, 823,
  \dodoi{10.1051/0004-6361:20065096}

\bibitem[{{Gregory} {et~al.}(1996){Gregory}, {Scott}, {Douglas}, \&
  {Condon}}]{1996ApJS..103..427G}
{Gregory}, P.~C., {Scott}, W.~K., {Douglas}, K., \& {Condon}, J.~J. 1996,
  \apjs, 103, 427, \dodoi{10.1086/192282}

\bibitem[{{Gu} {et~al.}(2022){Gu}, {Zhang}, {Gan}, {Zhang}, {Sun}, \&
  {Liang}}]{2022ApJ...927..221G}
{Gu}, Y., {Zhang}, H.-M., {Gan}, Y.-Y., {et~al.} 2022, \apj, 927, 221,
  \dodoi{10.3847/1538-4357/ac540e}

\bibitem[{{Healey} {et~al.}(2007){Healey}, {Romani}, {Taylor}, {Sadler},
  {Ricci}, {Murphy}, {Ulvestad}, \& {Winn}}]{2007ApJS..171...61H}
{Healey}, S.~E., {Romani}, R.~W., {Taylor}, G.~B., {et~al.} 2007, \apjs, 171,
  61, \dodoi{10.1086/513742}

\bibitem[{{Helfand} {et~al.}(2015){Helfand}, {White}, \&
  {Becker}}]{2015ApJ...801...26H}
{Helfand}, D.~J., {White}, R.~L., \& {Becker}, R.~H. 2015, \apj, 801, 26,
  \dodoi{10.1088/0004-637X/801/1/26}

\bibitem[{{Kalberla} {et~al.}(2005){Kalberla}, {Burton}, {Hartmann}, {Arnal},
  {Bajaja}, {Morras}, \& {P{\"o}ppel}}]{2005AA...440..775K}
{Kalberla}, P.~M.~W., {Burton}, W.~B., {Hartmann}, D., {et~al.} 2005, \aap,
  440, 775, \dodoi{10.1051/0004-6361:20041864}

\bibitem[{{Kiehlmann} {et~al.}(2024{\natexlab{a}}){Kiehlmann}, {Readhead},
  {O'Neill}, {Wilkinson}, {Lister}, {Liodakis}, {Bruzewski}, {Pavlidou},
  {Pearson}, {Sheldahl}, {Siemiginowska}, {Tassis}, \&
  {Taylor}}]{2024ApJ...961..241K}
{Kiehlmann}, S., {Readhead}, A.~C.~S., {O'Neill}, S., {et~al.}
  2024{\natexlab{a}}, \apj, 961, 241, \dodoi{10.3847/1538-4357/ad0cc2}

\bibitem[{{Kiehlmann} {et~al.}(2024{\natexlab{b}}){Kiehlmann}, {Lister},
  {Readhead}, {Liodakis}, {O'Neill}, {Pearson}, {Sheldahl}, {Siemiginowska},
  {Tassis}, {Taylor}, \& {Wilkinson}}]{2024ApJ...961..240K}
{Kiehlmann}, S., {Lister}, M.~L., {Readhead}, A.~C.~S., {et~al.}
  2024{\natexlab{b}}, \apj, 961, 240, \dodoi{10.3847/1538-4357/ad0c56}

\bibitem[{{Kino} \& {Asano}(2011)}]{2011MNRAS.412L..20K}
{Kino}, M., \& {Asano}, K. 2011, \mnras, 412, L20,
  \dodoi{10.1111/j.1745-3933.2010.00996.x}

\bibitem[{{Kosmaczewski} {et~al.}(2020){Kosmaczewski}, {Stawarz},
  {Siemiginowska}, {Cheung}, {Ostorero}, {Sobolewska}, {Kozie{\l}-Wierzbowska},
  {W{\'o}jtowicz}, \& {Marchenko}}]{2020ApJ...897..164K}
{Kosmaczewski}, E., {Stawarz}, {\L}., {Siemiginowska}, A., {et~al.} 2020, \apj,
  897, 164, \dodoi{10.3847/1538-4357/ab9b1f}

\bibitem[{{Lacy} {et~al.}(2020){Lacy}, {Baum}, {Chandler}, {Chatterjee},
  {Clarke}, {Deustua}, {English}, {Farnes}, {Gaensler}, {Gugliucci},
  {Hallinan}, {Kent}, {Kimball}, {Law}, {Lazio}, {Marvil}, {Mao}, {Medlin},
  {Mooley}, {Murphy}, {Myers}, {Osten}, {Richards}, {Rosolowsky}, {Rudnick},
  {Schinzel}, {Sivakoff}, {Sjouwerman}, {Taylor}, {White}, {Wrobel},
  {Andernach}, {Beasley}, {Berger}, {Bhatnager}, {Birkinshaw}, {Bower},
  {Brandt}, {Brown}, {Burke-Spolaor}, {Butler}, {Comerford}, {Demorest}, {Fu},
  {Giacintucci}, {Golap}, {G{\"u}th}, {Hales}, {Hiriart}, {Hodge}, {Horesh},
  {Ivezi{\'c}}, {Jarvis}, {Kamble}, {Kassim}, {Liu}, {Loinard}, {Lyons},
  {Masters}, {Mezcua}, {Moellenbrock}, {Mroczkowski}, {Nyland}, {O'Dea},
  {O'Sullivan}, {Peters}, {Radford}, {Rao}, {Robnett}, {Salcido}, {Shen},
  {Sobotka}, {Witz}, {Vaccari}, {van Weeren}, {Vargas}, {Williams}, \&
  {Yoon}}]{2020PASP..132c5001L}
{Lacy}, M., {Baum}, S.~A., {Chandler}, C.~J., {et~al.} 2020, \pasp, 132,
  035001, \dodoi{10.1088/1538-3873/ab63eb}

\bibitem[{{Lister} {et~al.}(2020){Lister}, {Homan}, {Kovalev}, {Mandal},
  {Pushkarev}, \& {Siemiginowska}}]{lister2020}
{Lister}, M.~L., {Homan}, D.~C., {Kovalev}, Y.~Y., {et~al.} 2020, \apj, 899,
  141, \dodoi{10.3847/1538-4357/aba18d}

\bibitem[{{Marchesi} {et~al.}(2018){Marchesi}, {Ajello}, {Marcotulli},
  {Comastri}, {Lanzuisi}, \& {Vignali}}]{2018ApJ...854...49M}
{Marchesi}, S., {Ajello}, M., {Marcotulli}, L., {et~al.} 2018, \apj, 854, 49,
  \dodoi{10.3847/1538-4357/aaa410}

\bibitem[{{Matake} \& {Fukazawa}(2023)}]{2023PASJ...75.1124M}
{Matake}, H., \& {Fukazawa}, Y. 2023, \pasj, 75, 1124,
  \dodoi{10.1093/pasj/psad060}

\bibitem[{{Mateos} {et~al.}(2012){Mateos}, {Alonso-Herrero}, {Carrera},
  {Blain}, {Watson}, {Barcons}, {Braito}, {Severgnini}, {Donley}, \&
  {Stern}}]{2012MNRAS.426.3271M}
{Mateos}, S., {Alonso-Herrero}, A., {Carrera}, F.~J., {et~al.} 2012, \mnras,
  426, 3271, \dodoi{10.1111/j.1365-2966.2012.21843.x}

\bibitem[{{Mattox} {et~al.}(1996){Mattox}, {Bertsch}, {Chiang}, {Dingus},
  {Digel}, {Esposito}, {Fierro}, {Hartman}, {Hunter}, {Kanbach}, {Kniffen},
  {Lin}, {Macomb}, {Mayer-Hasselwander}, {Michelson}, {von Montigny},
  {Mukherjee}, {Nolan}, {Ramanamurthy}, {Schneid}, {Sreekumar}, {Thompson}, \&
  {Willis}}]{1996ApJ...461..396M}
{Mattox}, J.~R., {Bertsch}, D.~L., {Chiang}, J., {et~al.} 1996, \apj, 461, 396,
  \dodoi{10.1086/177068}

\bibitem[{{Mauch} {et~al.}(2003){Mauch}, {Murphy}, {Buttery}, {Curran},
  {Hunstead}, {Piestrzynski}, {Robertson}, \& {Sadler}}]{2003MNRAS.342.1117M}
{Mauch}, T., {Murphy}, T., {Buttery}, H.~J., {et~al.} 2003, \mnras, 342, 1117,
  \dodoi{10.1046/j.1365-8711.2003.06605.x}

\bibitem[{{Migliori} {et~al.}(2016){Migliori}, {Siemiginowska}, {Sobolewska},
  {Loh}, {Corbel}, {Ostorero}, \& {Stawarz}}]{2016ApJ...821L..31M}
{Migliori}, G., {Siemiginowska}, A., {Sobolewska}, M., {et~al.} 2016, \apjl,
  821, L31, \dodoi{10.3847/2041-8205/821/2/L31}

\bibitem[{{Neugebauer} {et~al.}(1984){Neugebauer}, {Habing}, {van Duinen},
  {Aumann}, {Baud}, {Beichman}, {Beintema}, {Boggess}, {Clegg}, {de Jong},
  {Emerson}, {Gautier}, {Gillett}, {Harris}, {Hauser}, {Houck}, {Jennings},
  {Low}, {Marsden}, {Miley}, {Olnon}, {Pottasch}, {Raimond}, {Rowan-Robinson},
  {Soifer}, {Walker}, {Wesselius}, \& {Young}}]{1984ApJ...278L...1N}
{Neugebauer}, G., {Habing}, H.~J., {van Duinen}, R., {et~al.} 1984, \apjl, 278,
  L1, \dodoi{10.1086/184209}

\bibitem[{{O'Dea} \& {Saikia}(2021)}]{2021A&ARv..29....3O}
{O'Dea}, C.~P., \& {Saikia}, D.~J. 2021, \aapr, 29, 3,
  \dodoi{10.1007/s00159-021-00131-w}

\bibitem[{{Orienti}(2016)}]{2016AN....337....9O}
{Orienti}, M. 2016, Astronomische Nachrichten, 337, 9,
  \dodoi{10.1002/asna.201512257}

\bibitem[{{Petrov}(2021)}]{2021AJ....161...14P}
{Petrov}, L. 2021, \aj, 161, 14, \dodoi{10.3847/1538-3881/abc4e1}

\bibitem[{{Principe} {et~al.}(2021){Principe}, {Di Venere}, {Orienti},
  {Migliori}, {D'Ammando}, {Mazziotta}, \& {Giroletti}}]{2021MNRAS.507.4564P}
{Principe}, G., {Di Venere}, L., {Orienti}, M., {et~al.} 2021, \mnras, 507,
  4564, \dodoi{10.1093/mnras/stab2357}

\bibitem[{{Principe} {et~al.}(2020){Principe}, {Migliori}, {Johnson},
  {D'Ammando}, {Giroletti}, {Orienti}, {Stanghellini}, {Taylor}, {Torresi}, \&
  {Cheung}}]{Principe2020}
{Principe}, G., {Migliori}, G., {Johnson}, T.~J., {et~al.} 2020, \aap, 635,
  A185, \dodoi{10.1051/0004-6361/201937049}

\bibitem[{{Readhead} {et~al.}(1996{\natexlab{a}}){Readhead}, {Taylor}, {Xu},
  {Pearson}, {Wilkinson}, \& {Polatidis}}]{readhead1996}
{Readhead}, A.~C.~S., {Taylor}, G.~B., {Xu}, W., {et~al.} 1996{\natexlab{a}},
  \apj, 460, 612, \dodoi{10.1086/176996}

\bibitem[{{Readhead} {et~al.}(1996{\natexlab{b}}){Readhead}, {Taylor}, {Xu},
  {Pearson}, {Wilkinson}, \& {Polatidis}}]{1996ApJ...460..612R}
---. 1996{\natexlab{b}}, \apj, 460, 612, \dodoi{10.1086/176996}

\bibitem[{{Readhead} {et~al.}(2024){Readhead}, {Ravi}, {Blandford}, {Sullivan},
  {Somalwar}, {Begelman}, {Birkinshaw}, {Liodakis}, {Lister}, {Pearson},
  {Taylor}, {Wilkinson}, {Globus}, {Kiehlmann}, {Lawrence}, {Murphy},
  {O'Neill}, {Pavlidou}, {Sheldahl}, {Siemiginowska}, \&
  {Tassis}}]{2024ApJ...961..242R}
{Readhead}, A.~C.~S., {Ravi}, V., {Blandford}, R.~D., {et~al.} 2024, \apj, 961,
  242, \dodoi{10.3847/1538-4357/ad0c55}

\bibitem[{{Shimwell} {et~al.}(2022){Shimwell}, {Hardcastle}, {Tasse}, {Best},
  {R{\"o}ttgering}, {Williams}, {Botteon}, {Drabent}, {Mechev}, {Shulevski},
  {van Weeren}, {Bester}, {Br{\"u}ggen}, {Brunetti}, {Callingham}, {Chy{\.z}y},
  {Conway}, {Dijkema}, {Duncan}, {de Gasperin}, {Hale}, {Haverkorn}, {Hugo},
  {Jackson}, {Mevius}, {Miley}, {Morabito}, {Morganti}, {Offringa}, {Oonk},
  {Rafferty}, {Sabater}, {Smith}, {Schwarz}, {Smirnov}, {O'Sullivan},
  {Vedantham}, {White}, {Albert}, {Alegre}, {Asabere}, {Bacon}, {Bonafede},
  {Bonnassieux}, {Brienza}, {Bilicki}, {Bonato}, {Calistro Rivera}, {Cassano},
  {Cochrane}, {Croston}, {Cuciti}, {Dallacasa}, {Danezi}, {Dettmar}, {Di
  Gennaro}, {Edler}, {En{\ss}lin}, {Emig}, {Franzen}, {Garc{\'\i}a-Vergara},
  {Grange}, {G{\"u}rkan}, {Hajduk}, {Heald}, {Heesen}, {Hoang}, {Hoeft},
  {Horellou}, {Iacobelli}, {Jamrozy}, {Jeli{\'c}}, {Kondapally}, {Kukreti},
  {Kunert-Bajraszewska}, {Magliocchetti}, {Mahatma}, {Ma{\l}ek}, {Mandal},
  {Massaro}, {Meyer-Zhao}, {Mingo}, {Mostert}, {Nair}, {Nakoneczny},
  {Nikiel-Wroczy{\'n}ski}, {Orr{\'u}}, {Pajdosz-{\'S}mierciak}, {Pasini},
  {Prandoni}, {van Piggelen}, {Rajpurohit}, {Retana-Montenegro}, {Riseley},
  {Rowlinson}, {Saxena}, {Schrijvers}, {Sweijen}, {Siewert}, {Timmerman},
  {Vaccari}, {Vink}, {West}, {Wo{\l}owska}, {Zhang}, \&
  {Zheng}}]{2022A&A...659A...1S}
{Shimwell}, T.~W., {Hardcastle}, M.~J., {Tasse}, C., {et~al.} 2022, \aap, 659,
  A1, \dodoi{10.1051/0004-6361/202142484}

\bibitem[{{Skrutskie} {et~al.}(2006){Skrutskie}, {Cutri}, {Stiening},
  {Weinberg}, {Schneider}, {Carpenter}, {Beichman}, {Capps}, {Chester},
  {Elias}, {Huchra}, {Liebert}, {Lonsdale}, {Monet}, {Price}, {Seitzer},
  {Jarrett}, {Kirkpatrick}, {Gizis}, {Howard}, {Evans}, {Fowler}, {Fullmer},
  {Hurt}, {Light}, {Kopan}, {Marsh}, {McCallon}, {Tam}, {Van Dyk}, \&
  {Wheelock}}]{2006AJ....131.1163S}
{Skrutskie}, M.~F., {Cutri}, R.~M., {Stiening}, R., {et~al.} 2006, \aj, 131,
  1163, \dodoi{10.1086/498708}

\bibitem[{{Sobolewska} {et~al.}(2022){Sobolewska}, {Migliori}, {Ostorero},
  {Siemiginowska}, {Stawarz}, {Guainazzi}, \& {Hardcastle}}]{Sobolewska2022}
{Sobolewska}, M., {Migliori}, G., {Ostorero}, L., {et~al.} 2022, \apj, 941, 52,
  \dodoi{10.3847/1538-4357/ac98ba}

\bibitem[{{Stanghellini} {et~al.}(1993){Stanghellini}, {O'Dea}, {Baum}, \&
  {Laurikainen}}]{1993ApJS...88....1S}
{Stanghellini}, C., {O'Dea}, C.~P., {Baum}, S.~A., \& {Laurikainen}, E. 1993,
  \apjs, 88, 1, \dodoi{10.1086/191812}

\bibitem[{{Stawarz} {et~al.}(2008){Stawarz}, {Ostorero}, {Begelman},
  {Moderski}, {Kataoka}, \& {Wagner}}]{Stawarz2008}
{Stawarz}, {\L}., {Ostorero}, L., {Begelman}, M.~C., {et~al.} 2008, \apj, 680,
  911, \dodoi{10.1086/587781}

\bibitem[{{Stern} {et~al.}(2012){Stern}, {Assef}, {Benford}, {Blain}, {Cutri},
  {Dey}, {Eisenhardt}, {Griffith}, {Jarrett}, {Lake}, {Masci}, {Petty},
  {Stanford}, {Tsai}, {Wright}, {Yan}, {Harrison}, \&
  {Madsen}}]{2012ApJ...753...30S}
{Stern}, D., {Assef}, R.~J., {Benford}, D.~J., {et~al.} 2012, \apj, 753, 30,
  \dodoi{10.1088/0004-637X/753/1/30}

\bibitem[{{White}(1992)}]{1992PASA...10..140W}
{White}, G.~L. 1992, \pasa, 10, 140, \dodoi{10.1017/S1323358000019500}

\bibitem[{{Wilkinson} {et~al.}(1994){Wilkinson}, {Polatidis}, {Readhead}, {Xu},
  \& {Pearson}}]{Wilkinson1994}
{Wilkinson}, P.~N., {Polatidis}, A.~G., {Readhead}, A.~C.~S., {Xu}, W., \&
  {Pearson}, T.~J. 1994, \apjl, 432, L87, \dodoi{10.1086/187518}

\bibitem[{{Wood} {et~al.}(2017){Wood}, {Caputo}, {Charles}, {Di Mauro},
  {Magill}, \& {Jeremy Perkins for the Fermi-LAT
  Collaboration}}]{2017arXiv170709551W}
{Wood}, M., {Caputo}, R., {Charles}, E., {et~al.} 2017, ArXiv e-prints.
\newblock \doarXiv{1707.09551}

\bibitem[{{Wright} \& {Otrupcek}(1990)}]{1990PKS...C......0W}
{Wright}, A., \& {Otrupcek}, R. 1990, PKS Catalog (1990, 0

\bibitem[{{Wright} {et~al.}(1994){Wright}, {Griffith}, {Burke}, \&
  {Ekers}}]{1994ApJS...91..111W}
{Wright}, A.~E., {Griffith}, M.~R., {Burke}, B.~F., \& {Ekers}, R.~D. 1994,
  \apjs, 91, 111, \dodoi{10.1086/191939}

\bibitem[{{Wright} {et~al.}(2010){Wright}, {Eisenhardt}, {Mainzer}, {Ressler},
  {Cutri}, {Jarrett}, {Kirkpatrick}, {Padgett}, {McMillan}, {Skrutskie},
  {Stanford}, {Cohen}, {Walker}, {Mather}, {Leisawitz}, {Gautier}, {McLean},
  {Benford}, {Lonsdale}, {Blain}, {Mendez}, {Irace}, {Duval}, {Liu}, {Royer},
  {Heinrichsen}, {Howard}, {Shannon}, {Kendall}, {Walsh}, {Larsen}, {Cardon},
  {Schick}, {Schwalm}, {Abid}, {Fabinsky}, {Naes}, \&
  {Tsai}}]{2010AJ....140.1868W}
{Wright}, E.~L., {Eisenhardt}, P. R.~M., {Mainzer}, A.~K., {et~al.} 2010, \aj,
  140, 1868, \dodoi{10.1088/0004-6256/140/6/1868}

\bibitem[{{Yamamura} {et~al.}(2010){Yamamura}, {Makiuti}, {Ikeda}, {Fukuda},
  {Oyabu}, {Koga}, \& {White}}]{2010yCat.2298....0Y}
{Yamamura}, I., {Makiuti}, S., {Ikeda}, N., {et~al.} 2010, {VizieR Online Data
  Catalog: AKARI/FIS All-Sky Survey Point Source Catalogues (ISAS/JAXA, 2010)},
  VizieR On-line Data Catalog: II/298. Originally published in: ISAS/JAXA
  (2010)

\bibitem[{{Yershov}(2014)}]{2014Ap&SS.354...97Y}
{Yershov}, V.~N. 2014, \apss, 354, 97, \dodoi{10.1007/s10509-014-1944-5}

\end{thebibliography}
\bibliographystyle{aasjournal}



\end{document}